\documentclass[preprint]{aastex}
\usepackage{color}
\usepackage[dvipdfm,pdfstartview=FitH,CJKbookmarks=true,
bookmarksnumbered=true,bookmarksopen=true]{hyperref}

\newcommand{\tabincell}[2]{\begin{tabular}{@{}#1 @{}}#2\end{tabular}}

\begin{document}

\title{The Silicon and Calcium High-Velocity Features in Type Ia Supernovae from Early to Maximum Phases}
\author{Xulin Zhao\altaffilmark{1,2}, Xiaofeng Wang\altaffilmark{1}, Keiichi Maeda\altaffilmark{2,3}, Hanna Sai\altaffilmark{1}, Tianmeng Zhang\altaffilmark{5}, Jujia Zhang\altaffilmark{6,7}, Fang Huang\altaffilmark{1,4}, Liming Rui\altaffilmark{1}, Qi Zhou\altaffilmark{1}, Jun Mo\altaffilmark{1}}

\altaffiltext{1}{Center for Astrophysics, Department of Physics, Tsinghua University, Beijing, 100084, China: zhaoxl11@mails.tsinghua.edu.cn; wang\_xf@mail.tsinghua.edu.cn}
\altaffiltext{2}{Department of Astronomy, Kyoto University, Kyoto, 606-8502, Japan}
 \altaffiltext{3}{Kavli Institute for the Physics and Mathematics of the Universe (WPI), University of Tokyo, 5-1-5 Kashiwanoha, Kashiwa, Chiba 277-8583, Japan}
\altaffiltext{4}{Department of Astronomy, Beijing Normal University, Beijing 100875, China} 
\altaffiltext{5}{National Astronomical Observatories, Chinese Academy of Sciences, Beijing 100012, China}
\altaffiltext{6}{Yunnan Observatories, Chinese Academy of Sciences, Kunming 650216, China} 
\altaffiltext{7}{Key Laboratory for the Structure and Evolution of Celestial Objects, Chinese Academy of Sciences, Kunming 650216, China}

\begin{abstract}
The high-velocity features (HVFs) in optical spectra of type Ia supernovae (SNe~Ia) are examined with a large sample including very early-time spectra (e.g., t $<$ $-$7 days). Multiple Gaussian fits are applied to examine the HVFs and their evolutions, using constraints on expansion velocities for the same species (i.e., Si~II 5972 and Si~II 6355). We find that strong HVFs tend to appear in SNe~Ia with smaller decline rates (e.g., $\Delta$m$_{15}$(B)$\lesssim$1.4 mag), clarifying that the finding by \citet{ch14} for the Ca-HVFs in near-maximum-light spectra applies both to the Si-HVFs and Ca-HVFs in the earlier phase. The Si-HVFs seem to be more common in fast-expanding SNe Ia, which is different from the earlier result that the Ca-HVFs are associated with SNe Ia having slower Si~II~6355 velocities at maximum light (i.e., $V^{Si}_{max}$). Moreover, SNe Ia with both stronger HVFs at early phases and larger $V^{Si}_{max}$ are found to have noticeably redder B$-$V colors and occur preferentially in the inner regions of their host galaxies, while those with stronger HVFs but smaller $V^{Si}_{max}$ show opposite tendencies, suggesting that these two subclasses have different explosion environments and their HVFs may have different origins. We further examine the relationships between the absorption features of Si~II~6355 and Ca~II~IR lines, and find that their photospheric components are well correlated in velocity and strength but the corresponding HVFs show larger scatter. These results cannot be explained with ionization and/or thermal processes alone, and different mechanisms are required for the creation of HVF-forming region in SNe Ia.
\end{abstract}


\section{Introduction\label{S_introduction}}

Type Ia supernovae (SNe Ia) represent one of the best tools measuring extragalactic distances, leading to the discovery of cosmic acceleration \citep{ri98a,pe99}. Despite great efforts to improve the understanding of SN Ia explosions, the nature of their progenitors has not yet been clarified \citep[e.g.,][]{kh91,hi00,va05,mae10,ka10,wb12,bl13,wa13,mag14,mao14}. There are two competing scenarios: the single degenerate scenario \citep[SD,][]{wh73,nom82b} which consists of a CO white dwarf (WD) and a non-degenerate companion star; and the double degenerate scenario which consists of two WDs \citep[DD,][]{ib84,we84}.

Constraints from the observations of nearby SN 2011fe in M101 \citep{nu11b} and the analyses of some supernova remnants of type Ia SNe such as SN 1006 and SNR 0509-67.5 in LMC favor the DD scenario because of the non-detection of a companion star with a luminosity greater than a few percent of the sun \citep{li11, blo12, bro12, her12, sch12}. On the other hand, there are also evidences for the presence of circumstellar material (CSM) around some SNe~Ia, which is in favor of the SD scenario \citep{ham03,ald06,pa07,st11,di12,mag13,sil13}. A recent study indicates that the spectral diversity observed in some SNe Ia (i.e., the scatter in the Si~II velocity) is correlated with the environments at the SN sites, further suggesting that SNe Ia may come from multiple progenitor systems \citep{wa13}.

Spectral features can provide clues to diagnose the origins of their observed diversities, including velocity and strength of the lines of different species, their high-velocity features (HVFs), and appearances of some rare absorption features such as unburned carbon \citep{pa11,tho11,fo12,si12d,hsi15}, variable sodium lines \citep{pa07}, or even H and/or He lines. To describe the spectroscopic
diversity of SNe Ia, different classification schemes have been proposed. \citet{be05} divided SNe Ia into `Faint' (i.e., SN
1991bg-like events), `high-velocity gradient (HVG)', and `low-velocity gradient (LVG)' subtypes. \citet{br06,br09} gave a
classification with subgroups of `core-normal (CN)', `broad-lined (BL)', `cool', and `shallow silicon (SS)'. \citet{wa09a} proposed that most of SNe Ia can be categorized into `Normal-Velocity (NV)' and `High-Velocity (HV)' subclasses in light of the photospheric
velocities measured from the Si~II$\lambda$6355 absorption near the maximum light. The above classification schemes show some overlaps between their subgroups, suggesting that some observed diversities may have similar origins. In general, the HV SNe Ia tend to have red B $-$ V colors around the maximum light \citep{wa09a, mae11, fol12} and their spectra do not show prominent signatures of carbon as compared to the NV counterparts \citep{fo12,si12d}.

Besides the diagnostics from the photospheric velocity and its evolution, the absorption feature appearing at higher velocities
(above the photosphere) can be used to classify subtypes of SNe Ia. Such HVFs are known to be seen in Ca~II IR triplet lines and
sometimes in Si~II$\lambda$6355 lines in some SNe Ia, for example the HVFs reported for SN 1999ee \citep{ma05b}, SN 2005cf \citep{ga07,wa09b}, SN 2009ig \citep{mar13}, and SN 2012fr \citep{ch13,zh14}. However, a physical origin of these HVFs is still obscured.

An increasing attention has been drawn for the HVFs in SNe Ia, and recent studies suggest correlations between different observables
such as the light-curve decline rate (i.e., $\Delta m_{15}(B)$), the peak B$-$V color, the velocity of Si~II~6355 around
maximum light ($V^{Si}_{max}$), and the properties of the host galaxies \citep[e.g., ][]{mag12,ch14,mag14,pa15,sil15}. These studies focused
primarily on the HVFs of Ca II NIR triplet. On the other hand, the HVF of Si~II~6355 has been rarely studied, except for a few
objects for which very early-time spectra are available (i.e., SN 2005cf, SN 2009ig, and SN 2012fr); this feature is relatively weak
and disappears quickly after the explosion. In this work, the HVFs are examined for both Si~II and Ca~II absorptions. This paper is
organized as follows. \S\ref{S2} presents the data sample and the measurements of the HVFs. In \S\ref{S3}, we examined the possible correlations
between the detected HVFs and other observables. The possible origins of the detached HVFs are discussed in \S\ref{S_origin}. The paper is closed in
\S\ref{S_conclusion} with conclusions.

\section{Datasets and Measurements of the Parameters\label{S2}}
\subsection{Sample}\label{S2.1}

This work performs a systematic search of the HVFs of Si~II and Ca~II in SNe Ia, with an attempt to examine a possible correlation
between them and to establish a link between the HVFs and other observables. This requires a sample of SNe Ia for which relatively early spectroscopic and well-sampled photometric observations are both available.

The spectral data used in our analysis are primarily from the CfA supernova program \citep{mat08,bl12}, the Berkeley supernova program
\citep{si12a,si12b}, Carnegie supernova project \citep[CSP,][]{fo13}, and our own database (Rui et al. in prep.). To secure accurate measurements of the HVFs, the spectra with low signal-to-noise (S/N) ratios (i.e., S/N $\lesssim$ 20 for the absorption features of Si~II~6355 and S/N $\lesssim$ 10 for the absorptions of Ca II IR triplet) are not included in our analysis. As summarized in Tables 1 and 2, our sample consists of a total of 316 early-time spectra with t $\leq$ $-7$ days. Among this sample, 76 spectra cover the wavelength extending beyond 9000\AA. These early-time spectra enable the studies of the HVFs of Si~II~6355 and Ca~II~IR triplet for 107 SNe Ia (sample A) and 46 SNe Ia (sample B, a subsample of sample A), respectively. In addition, we also collected a sample of 208 near-maximum-light spectra (i.e., $-3 \leq t \leq +3$ days) for 138 SNe Ia. This sample is used to investigate correlations of line velocities of Si~II~5972, Si~II~6355, and Ca~II~IR triplet. We did not attempt to measure the HVFs of Ca II H\&K, as they are usually blended with the Si~II~3850 absorption.

The sources of the light curves are the Harvard CfA SN group \citep{ri99,jh06,hi09,hi12}, the CSP \citep{con10, str11}, the Lick Observatory Supernova Search \citep[LOSS,][]{ga10}, and our database. We used SALT2 \citep{gu07} to obtain the peak magnitudes, color parameters, and stretch factors for SNe Ia. The stretch factor is then converted into the decline-rate parameter $\Delta$m$_{15}$(B) using an empirical relation between them \citep[i.e.,][]{gu07}. When multiple sources of light curves are available for an SN, we calculate the final $\Delta$m$_{15}$(B) and its uncertainty as the weighted mean of the values obtained by the SALT2 fitting.

\subsection{Measuring the HVFs}\label{S2.2}
All the spectra used in our analysis were first corrected for the redshifts of the host galaxies. Before measuring the absorption
features, the spectra were first smoothed over a 20-50\AA\ range with a locally weighted linear regression method to avoid detecting false signals, i.e., dips of the noise spikes in the data as a local absorption minimum. Then the spectra were normalized by the continuum. Following \citet{ch14}, we define the pseudo-continuum as a straight line connecting the line wings on both sides of the absorption feature.

In our analysis, a multiple Gaussian function was applied to the absorption features of Si~II and Ca~II lines that were normalized to the pseudo-continuum. The fitting function is described as follows:
\begin{equation}\label{1}
f(\lambda)=A_1
exp(-\frac{(\lambda_1-\lambda_1^{rest})^2}{2\sigma_1^2})+A_2
exp(-\frac{(\lambda_2-\lambda_2^{rest})^2}{2\sigma_2^2})+ \ldots+A_n
exp(-\frac{(\lambda_n-\lambda_n^{rest})^2}{2\sigma_n^2})
\end{equation}

\begin{equation}\label{2}
pEW=\sqrt{2\pi} A_1 \sigma_1 + \sqrt{2\pi} A_2 \sigma_2 + \ldots +
\sqrt{2\pi} A_n \sigma_n
\end{equation}

\begin{figure*}
\epsscale{.95} \plotone{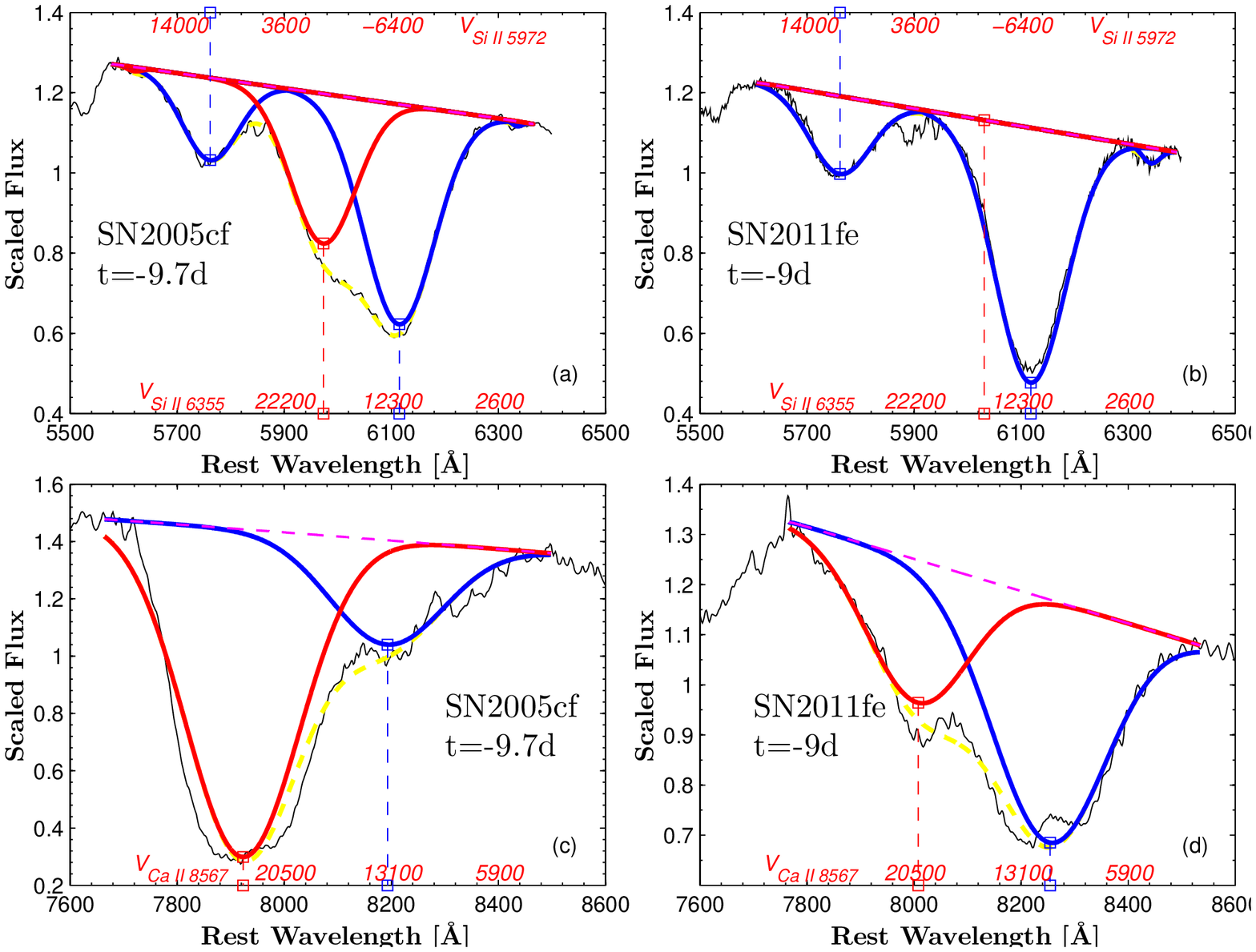} \caption{\label{FigFits} Gaussian fits to the absorptions of Si~II 6355 \AA~and Ca~II~IR lines in
t$\sim$ $-$10 day spectra of SN 2005cf and SN 2011fe. The blue lines show the fit to the photospheric components, and red lines show the
fit to the HVF components. The yellow dashed lines represent the combined fits. The weak absorption on the right side of Si~II 6355
represents the C~II~6580. The blue dashed line marks the position of absorption minimum of the photospheric component, and the red dashed
line shows the position of the absorption minimum of the high-velocity component.}
\end{figure*}

In Eq.(\ref{1}), subscripts `1', `2', and `n' denote different components formed at different velocities. `$A_{\rm i}$' represents the strength of the $i-$th absorption component, `$\sigma_{i}$' represents the full-width-at-half-maximum (FWHM) of the absorption and $\lambda_{rest}$ represents the rest-frame wavelength. Figure\ref{FigFits} shows fits to the early-phase spectra of two representative objects SN 2005cf and SN 2011fe.

A four-component gaussian function is used to fit the absorption features of Si~II~5972/6355 complex, which determines the photospheric and HVF
components of Si~II~6355 as well as the photospheric components of Si~II~5972 and C~II~6580 absorptions, respectively. Inclusion of the latter two components in the fit is due to that these two lines may affect the determinations of the line wings of Si~II~6355. As the HVF of Si~II~5972 is not identifiable in most SNe Ia \footnote{The HVF of Si~II~5972 may exist in the early spectra of some SNe Ia such as SN 2004dt where a five-component gaussian fit has to be used.}, the velocity inferred from absorption minimum of this line profile can thus serve as a good indicator of photospheric velocity of Si~II lines (see Figure 3 and discussions below). Although the C~II~6580 absorption is usually weak in most SNe Ia, it can be as strong as the Si~II~6355 absorption in some cases, i.e., SN 2006gz and SN 2013dy \citep{hi07, zh13}. For the fit to the Ca~II~IR triplet, we use the optically thick limit \citep[as in][]{ch14,mag14,pa15} to assume that the relative absorption strengths of the three lines are equal. This simply results in a single absorption component with a mean wavelength at $\sim$8567\AA\ for the Ca~II~IR triplet. We thus fit the observed line profile of Ca II IR absorptions with a two-component gaussian function, with one for the photospheric component and the other for the HVF in the velocity space (see Figure 1). This is somewhat inexact but not an unacceptable approximation as the velocity width of the line profiles generally exceeds that expected for the separation between Ca~II~8662 and Ca~II~8498 (which is about 5700 km s$^{-1}$) for most of our sample.

\begin{figure*}
\epsscale{.95} \plotone{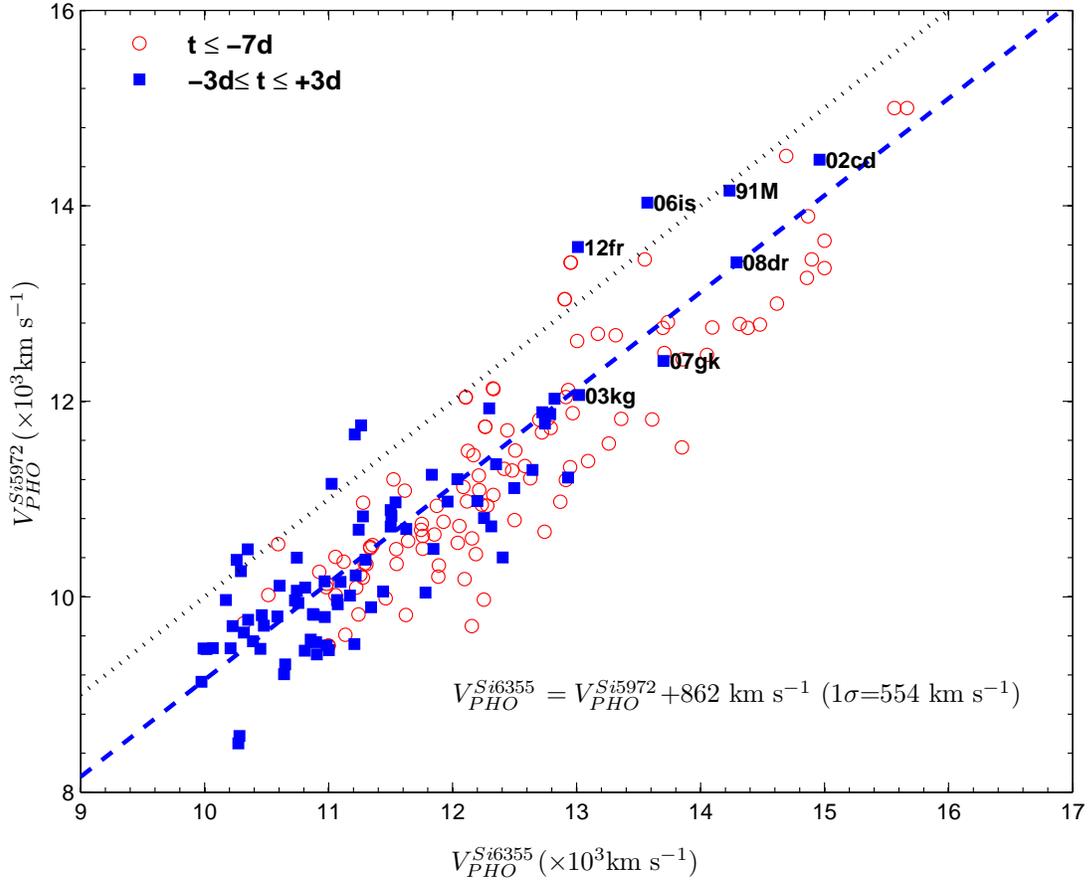}
\caption{\label{FigSiSi} The velocity relation between Si~II~5972 and Si~II~6355 measured from the near-maximum-light spectra (filled squares). Overplotted are the measurements of the corresponding velocities of these two Si~II lines from the early-time spectra (open circles). The blue dashed line shows the best linear fit to their correlations, while the dotted line marks the position where Si~II 5972 and Si~II 6355 have the
same velocity.}
\end{figure*}
\begin{figure*}
\epsscale{.95} \plotone{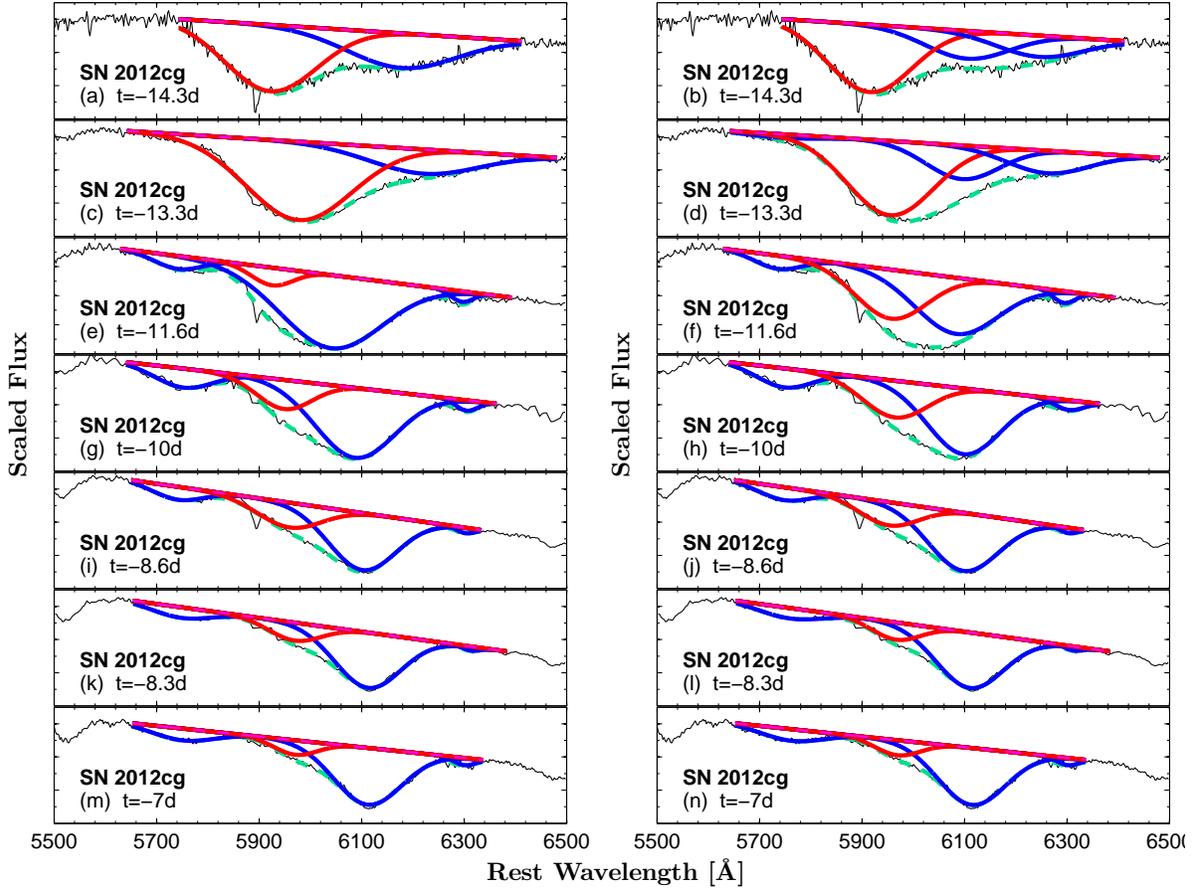} \caption{\label{FigPLot_12cg}
Gaussian fit to the early-time spectra of SN 2012cg. Left panels show the fit by assuming that the photospheric component of Si~II~6355 has a velocity close to that inferred from the absorption minimum. Right panels show the fit with constraints from the relation of the photospheric velocity between Si II 6355 and Si II 5972, as suggested by Fig.\ref{FigSiSi}. The blue lines show the fit to the photospheric components, and red lines show the fit to the HVF components. The green dashed lines represent
the combined fits.}
\end{figure*}

Relationships between velocities of different lines of the same species can be taken as an advantage to narrow down the parameter
space in the fitting \citep{ch14,pa15}. To avoid the possible contamination of the HVFs in our analysis, the near-maximum-light (i.e., at $-$3$<$ t $<$ +3 days) spectra are used to determine the velocity relation between Si~II~5972 and Si~II~6355. As shown in Fig. \ref{FigSiSi}, the photospheric velocity of Si II~5972 is found to be systematically lower than that of Si~II~6355 by 860$\pm$550 km s$^{-1}$ at around the maximum light, which seems to hold for the corresponding Si~II velocities measured at early times (e.g., t $<$ $-$ 7 days). Taking into consideration such a velocity relation and a 2-$\sigma$ uncertainty in fitting the Si~II~6355 absorptions, we can better separate the possible HVFs from the photospheric components, as shown in Figure \ref{FigPLot_12cg} for SN 2012cg. This velocity relation between Si~II~6355 and Si~II~5972 is thus employed in the fit to measure the photospheric velocity of Si~II~6355 for our sample with early-time spectra (e.g., sample A). Moreover, the measurements of photospheric velocity of Si~II~6355 near the maximum light are used to set a rough constraint on that of Ca~II~IR triplet, e.g., $V^{Ca}_{PHO}$ = $V^{Si}_{PHO}$$\pm$2,000 km s$^{-1}$. For the fit to the early-time Ca~II~IR triplet, such a velocity constraint may not be necessary because the HVFs of Ca~II~IR absorptions have remarkably high velocities at early times and are distinctly separated from the corresponding photospheric components.

Uncertainty in the measurement of the absorption velocity is estimated to be about 300 km s$^{-1}$ \citep[see also ][]{ch14}, depending primarily on the resolution and S/N ratio of the spectra, which correspond to an uncertainty of about 3\% for the velocity measurement. The uncertainties caused by the imperfect definition of the continuum and the fitting procedure are not included because of the difficulties in quantifying these effects. The uncertainty in the pseudo-equivalent width (pEW) of the absorption feature is larger, which depends mostly on the velocity measurements. In this paper, we assume an uncertainty of 5\% for the measurement of the pEW. Reddening should not have a significant effect, because the absorption features were normalized by the continuum.

\subsection{Subclassifications of the HVFs}\label{S2.3}

\begin{figure*}
\epsscale{.95} \plotone{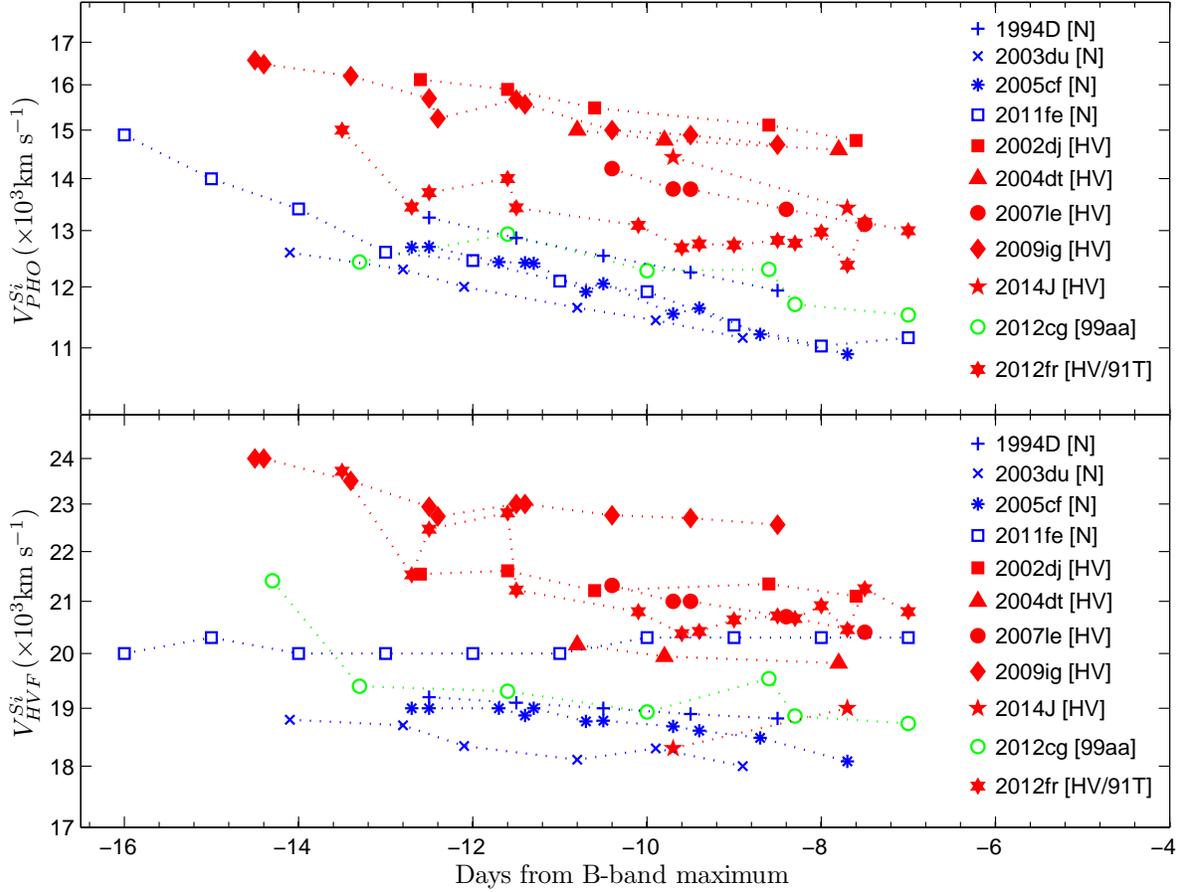} \caption{\label{FigVevo} Upper panel: evolution of the expansion velocity of the photospheric component of Si~II~6355 for some well-observed SNe Ia. Lower panel: evolution of the expansion velocity of the high-velocity features (HVFs). The `N', `HV', and `91T/99aa' listed in the brackets after each supernova represent the subclasses of the 'normal-velocity', 'high-velocity', and spectroscopically peculiar objects like SN 1991T/1999aa, respectively, following the classification scheme proposed by \citet{wa09a}. The sources of these spectra are listed in Table \ref{T_par}.}\end{figure*}

In Figure \ref{FigVevo}, we show the phase evolution of Si~II velocity for some well-observed SNe Ia with prominent HVFs at early phases, such as SNe 1994D, 2002dj, 2003du, 2005cf, 2007le, 2009ig, 2011fe, 2012cg, and 2012fr. Among these objects, SN 2002dj, SN 2007le, and SN 2009ig belong to the HV subclass, while SN 1994D, SN 2003du, SN 2005cf, and SN 2011fe can be put into the NV subclass according to classification scheme proposed by \citet{wa09a}. SN 2012cg and SN 2012fr show some similarities to the SN 1991T/1999aa-like subclass \citep{si12c, zh14}. For these sample, one can see that the photospheric components have expansion velocities ranging from $\lesssim$ 16,000 km s$^{-1}$ at very early phase (e.g., at t $<$-14 days) to $\sim$11,000 km s$^{-1}$ at t$\sim-$7 days; while the HVFs have much higher values, with the velocity varying in a range from $\sim$24,000 km s$^{-1}$ to $\sim$ 18,000 km s$^{-1}$. This indicates that the HVFs are formed at outer layers with a velocity $\gtrsim$5,000 km s$^{-1}$ above the photosphere.

\begin{figure*}
\epsscale{.95} \plotone{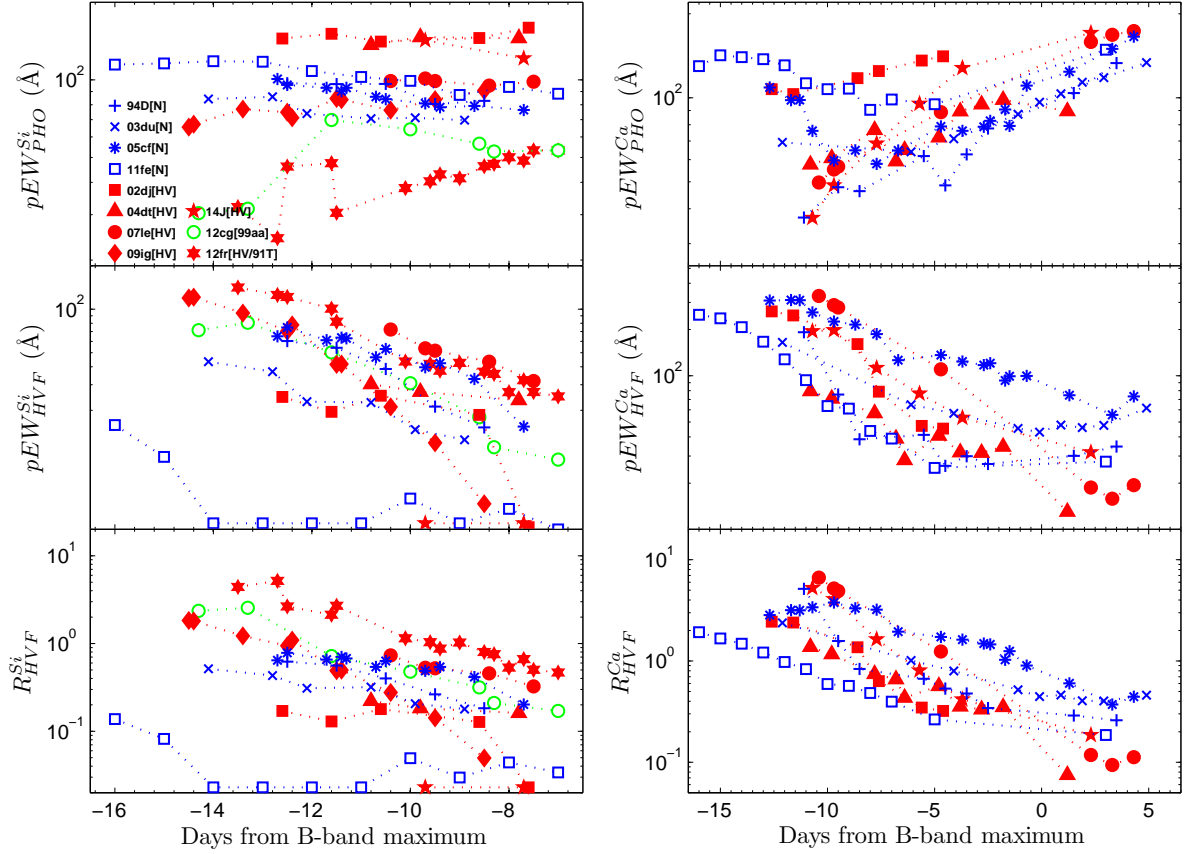}
\caption{\label{FigWevo} Left panels: evolution of the pseudo-equivalent width (pEW) of photospheric component (PHO, top panel), high-velocity feature (HVF, middle panel), and HVF/PHO ratio (bottom panel) of Si~II~6355 absorption. Right panels: the same evolution of the pEW but for the absorptions of Ca~II~IR triplet.}
\end{figure*}

As shown in Figure \ref{FigWevo} for some well-observed SNe Ia, the photospheric component of Si~II~6355 and Ca~II~IR triplet show a steady increase in the absorption strength when approaching the maximum light. On the other hand, their HVFs show a rapid decrease after the explosion. Such HVFs become very weak in Si~II at about one week before the maximum light, but they are detectable in Ca~II~IR near the maximum light or even at a few days after that. This explains why the HVFs are rarely seen in Si~II but are more commonly in Ca~II. Note that the HVFs of the SNe Ia with higher Si~II velocities at maximum light (i.e., $V^{Si}_{max}$) tend to decrease in strength at a rate faster than those with lower velocities. Following \citet{ch14}, we quantify the strength of the HVFs using the ratio of the pEW of the HVF to that of the photospheric component (PHO), e.g., R$_{HVF}$ = pEW(HVF)/pEW(PHO). The bottom panels of Figure \ref{FigWevo} show the temporal evolution of R$_{HVF}$. One can see that there is a large scatter in R$_{HVF}$ and its evolution in the earlier phases (see also Figure \ref{FigdRsidt}); and comparison of R$_{HVF}$ between different SNe Ia is thus sensitive to the supernova phase. To alleviate the evolutionary effect on the comparison, we normalize the R$_{HVF}$ measured at early phases (e.g., t $<$ $-$ 7 days) to an approximated value at the same phase (i.e., at t$\sim$ $-$ 8 days) by using the well-observed sample as templates for interpolations in the following analyses.

\begin{figure*}
\epsscale{.95} \plotone{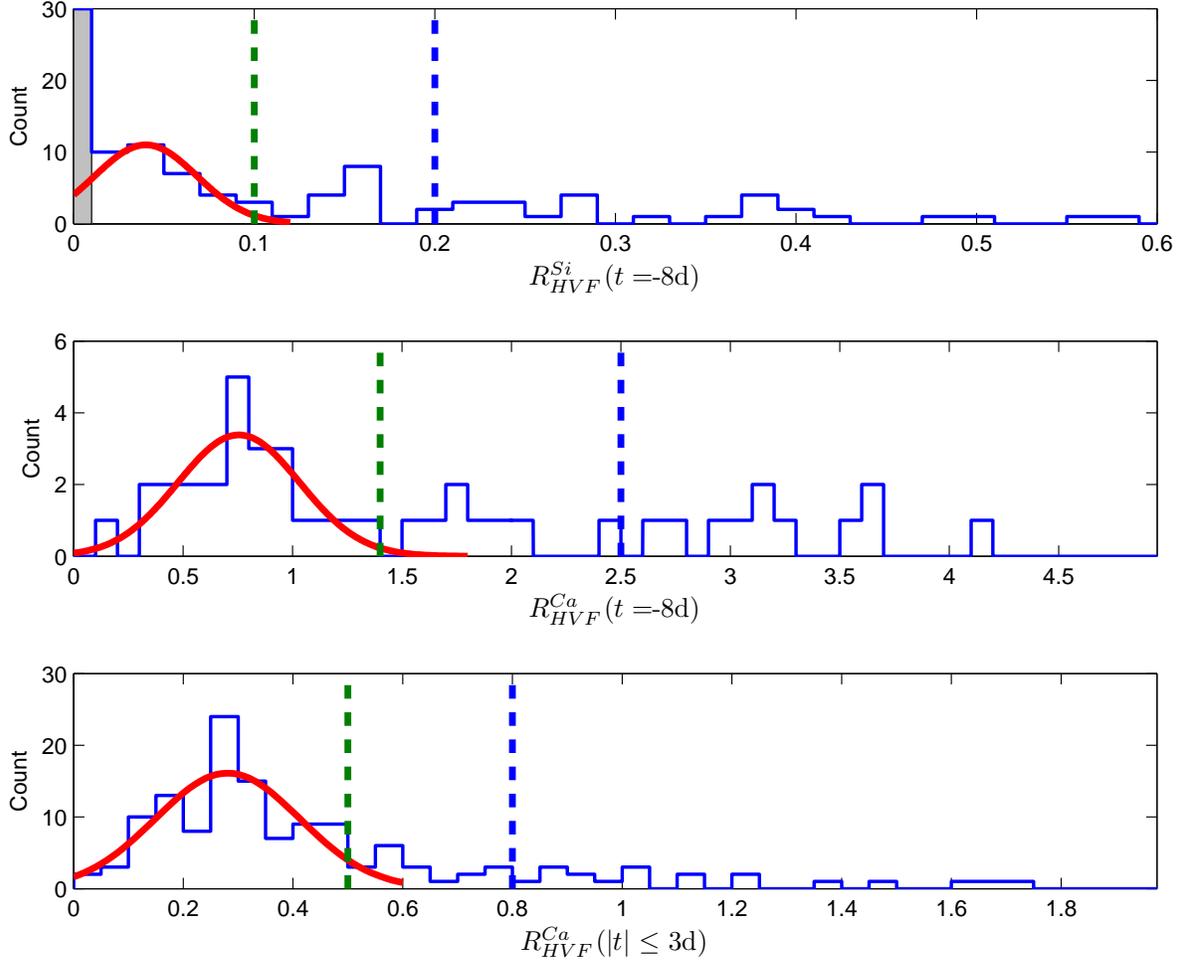}
\caption{\label{FigRhvf_dis} The histogram distributions of the ratio of the pEW of the high-velocity feature to that of the
photospheric component (which is defined as R$_{HVF}$). The upper, middle, and bottom panels represent the R$_{HVF}$ distribution for Si~II~6355 at $t\sim$ $-$8 days, the Ca~II~IR at t$\sim$$-$8 days, and the Ca~II~IR at t$\sim$0 days, respectively. A Gaussian function is used to fit the distribution of the HVF-weak SNe Ia in each panel. Two vertical lines mark the boundaries to distinguish the HVF-weak, in-between, and HVF-strong
subclasses.}
\end{figure*}

Figure \ref{FigRhvf_dis} shows the histogram distributions of R$_{HVF}$ from different samples. For Ca~II~IR triplet, the R$_{HVF}$ distribution can be decomposed into a gaussian component and a long tail, and the gaussian components are found to have peaks at 0.74$\pm$0.36 (1 $\sigma$) and 0.27$\pm$0.12 (1 $\sigma$) for the measurements made with t$\approx$ $-$ 8 day and t $\approx$ 0 day spectra, respectively. Such a trend does not seem to hold for the R$_{HVF}$ distribution of early-time Si~II~6355. Note, however, that the prominent peak of R$^{Si}_{HVF}$ at around 0 (see top panel of Figure \ref{FigRhvf_dis}) is an artifact of inaccurate measurements because it is usually assumed to be 0 when the Si-HVFs cannot be well identified. Neglecting this abnormal peak at R$^{Si}_{HVF}$ $\thickapprox$0, a gaussian function (with a peak of 0.035 and 1-$\sigma$ error of 0.022) can be used to describe the R$^{Si}_{HVF}$ distribution of Si~II lines for most SNe Ia. Therefore, we define the Si~II HVF-weak subclass of SNe Ia as those having R$^{Si}_{HVF}$ $<$ 0.1 at t$\approx -$8 days, while those R$^{Si}_{HVF}$ $\gtrsim$ 0.2 are defined as Si~II HVF-strong subclass. This finally yields 29 HVF-strong and 64 HVF-weak SNe Ia, which is about 27.1\% and 59.8\% of sample A (107 SNe Ia in total). Considering the measurement errors, the rest 14 (13.1\%) SNe Ia cannot be clearly classified and may lie between the HVF-strong and HVF-weak subclasses.

The HVFs usually appear stronger in the Ca~II IR triplet compared to those in the Si~II absorptions. In this paper, the HVFs of Ca are thus investigated with two samples, including 46 SNe Ia with early-time spectra (sample B) and 138 SNe Ia with near-maximum-light spectra (sample C). The larger samples collected in this study allow us to set tighter criteria for selecting the HVF-strong and HVF-weak SNe than previous studies. Based on the distribution of R$^{Ca}_{HVF}$ shown in Figure \ref{FigRhvf_dis}, we set the upper limit of the HVF-weak subgroup for Sample B as R$^{Ca}_{HVF}$ $\sim$ 1.4 and the lower limit of the HVF-strong subgroup as R$^{Ca}_{HVF}$ $\sim$ 2.5. To better distinguish the above two subgroups, we also defined the in-between subgroup as those with 1.4 $<$ R$^{Ca}_{HVF}$ $<$ 2.5. With these criteria, we finally found that 15 SNe Ia can be put into the HVF-strong subclass (32.6\%) and 23 ones can be put into the HVF-weak subclass (50.0\%), while there are 8 objects locating between them (17.4\%). For sample C (with $-$3 days $<$ t $<$ +3 days spectra), the criteria set for the HVF-strong and HVF-weak subsets are R$^{Ca}_{HVF}$ $\gtrsim$ 0.8 and R$^{Ca}_{HVF}$ $<$ 0.5, respectively. This leads to 22 SNe with prominent HVFs of Ca~II~IR triplet (16.0\%), 98 SNe with weak HVFs (71.0\%), and 18 in-between objects (13.0\%). Definitions of samples A, B, and C and the subclassifications of HVF-strong, HVF-weak, and In-between subclasses are listed in Table \ref{T_criterion}. The detailed fitting results of the absorption features of Si~II~6355/5972 and Ca~II~IR triplet from different samples are presented in Tables 3-5.

To examine the differences of the HVF-strong SNe Ia identified with different samples (i.e., early-time and near-maximum-light spectra) and selection criteria (i.e., Si~II and Ca~II lines), we compared the HVF-strong SNe Ia from sample A, sample B, and sample C. For samples A and B set up with the early-time spectra, there are 45 SNe in common. Among the 19 overlapping SNe with strong Si-HVFs (i.e., R$_{HVF}$ $\gtrsim$ 0.2), 15 were found to show prominent Ca-HVFs (i.e., R$^{Ca}_{HVF}$$\geq$2.5) and the other 4 can be put into the in-between subset (i.e., 1.4$<$ R$^{Ca}_{HVF}$ $<$ 2.5). However, among the 11 SNe Ia showing strong early-time Si-HVFs from samples A and C, only 4 show strong Ca-HVFs at around the maximum light. Such a discrepancy shows that the HVFs may occur in Si and Ca simultaneously at early phases for most SNe Ia but they show different temporal evolution after that. This implies that, when using HVFs at different phases to study their correlations with other observables, we may find different tendencies. In principle, it should be better to take into account the R$^{Si}_{HVF}$ and R$^{Ca}_{HVF}$ together in the classifications rather than consider them separately as we and other studies did. However, this would need a larger sample with early spectra and wavelength coverage extending to about 9,000\AA, which is only available for a few SNe Ia.

\section{Properties of SNe Ia with Strong and Weak HVFs\label{S3}}
SNe Ia are known to exhibit diverse photometric properties that are related to their spectroscopic diversities. For example, the 91T-like SNe Ia have relatively high luminosity, which are characterized by weak Si~II absorption and strong Fe~II and Fe III absorptions \citep{fil92a,phi92}; the 91bg-like SNe Ia have low luminosity, which displays strong Si~II 5972 absorption and Ti~II absorption near 4,000\AA\ \citep{fil92b}. Within `normal' SNe Ia, the SNe with faster expansion velocities are found to have redder $B - V$ colors at the maximum light than those with lower velocities \citep{wa09a}. Based on the detections of HVFs of Si~II and Ca~II lines in early-phase spectra of SNe Ia, we could also examine whether there are correlations with the observables such as $\Delta$m$_{15}$(B), peak B $-$ V color, Si~II velocity $V^{Si}_{max}$, and explosion environments etc. Figure  \ref{FigRhvf_dm}-\ref{FigHVF_ks_gal} show details about how these properties are related with the HVFs.

\subsection{Light-Curve Decline Rate $\Delta \rm m_{15}(B)$}\label{S_dm15}
\begin{figure*}
\epsscale{.95} \plotone{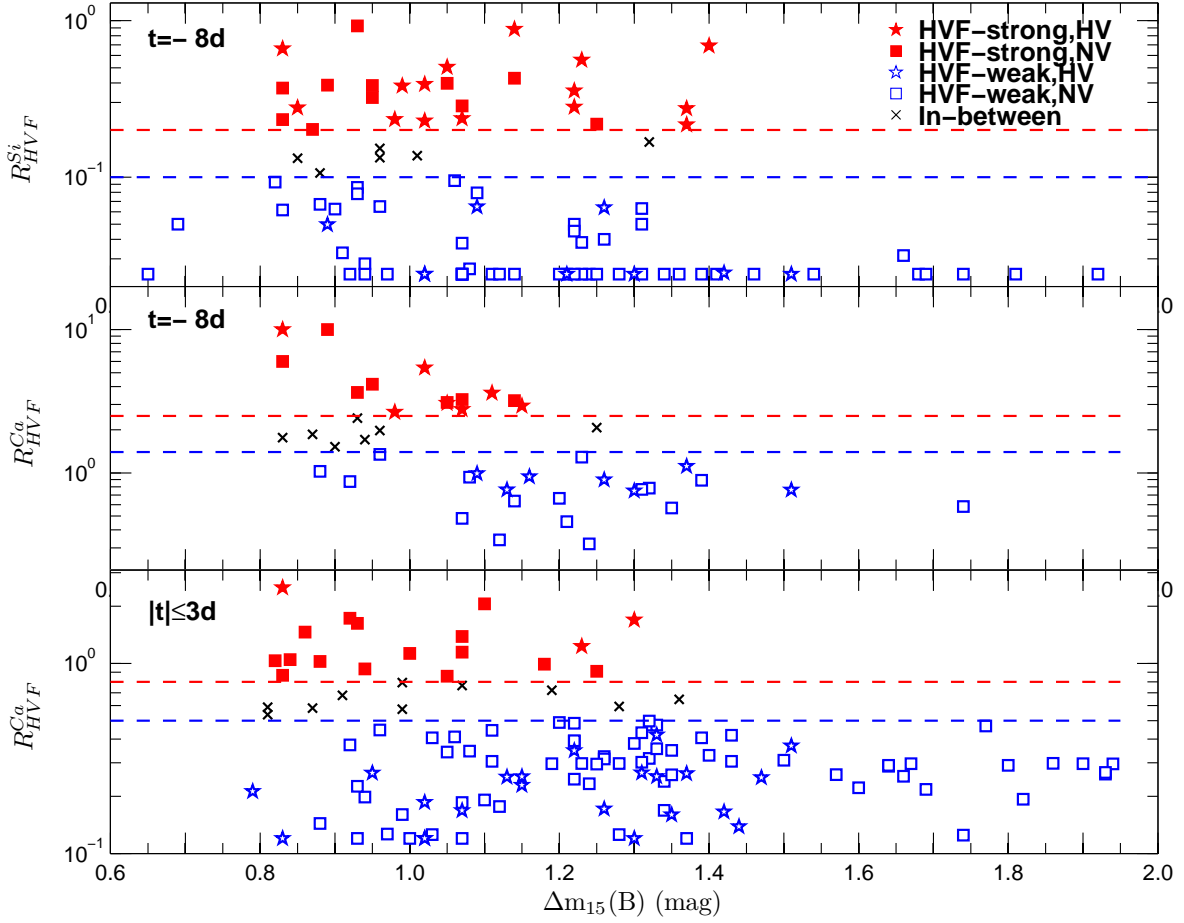}
\caption{\label{FigRhvf_dm} R$_{HVF}$ as a function of the light-curve decline rate $\Delta$m$_{15}$(B). Top panel: the distribution obtained from t$\sim$$-$8 day Si~II~6355 absorption of sample A. Middle panel: the case for the Ca~II~IR triplet in the t$\sim$$-$8 day spectra. Bottom panel: the case for the Ca~II~IR triplet in the t$\sim$0 day spectra. The stars represent the SN Ia sample with maximum-light $V^{Si}_{max}$ $\gtrsim$ 12,000 km s$^{-1}$, while the squares show the sample with $V^{Si}_{max}$ $<$ 12,000 km s$^{-1}$. The red dashed line is used to mark the boundary to divide the sample into HVF-strong and in-between subsamples, and the blue dashed line marks the boundary between the HVF-weak and in-between subsamples.}
\end{figure*}

The top panel of Figure \ref{FigRhvf_dm} shows the scatter plot between the R$^{Si}_{HVF}$ and the light-curve decline rate $\Delta$m$_{15}$(B). One can see that the strong HVFs of Si~II are rarely detected in SNe Ia with larger decline rates. For the 29 SNe Ia with R$^{Si}_{HVF}$$\gtrsim$0.2 at t$\approx$ $-$8 days, all are found to have $\Delta$m$_{15}$(B) $\lesssim$ 1.40 mag. The mean values of $\Delta$m$_{15}$(B) are found to be 1.06$\pm$0.17 mag and 1.20$\pm$0.27 mag, respectively, for the SNe Ia with R$^{Si}_{HVF}$$\gtrsim$0.2 and those with R$^{Si}_{HVF}$ $<$ 0.1. A K-S test indicates that there is a probability of only 3.5\% that these two subgroups of SNe Ia come from the same parent population (see Figure \ref{FigHVF_ks} and Table \ref{T_ks}).

Such a discrepancy in $\Delta m_{15}$(B) distribution can be also seen in the HVFs of Ca~II~IR triplet, as shown in the middle and bottom panels of Figure \ref{FigRhvf_dm}. For the subsamples classified by the early-time spectra of sample B, the HVF-strong SNe Ia (i.e., R$^{Ca}_{HVF}$$\gtrsim$2.5) have light-curve decline rates that are on average smaller than the HVF-weak SNe Ia (i.e., R$^{Ca}_{HVF}$ $<$ 1.4), with the mean value being as 0.99$\pm$0.12 mag and 1.19$\pm$0.20 mag, respectively. The K-S test shows that there is a probability of about 0.1\% that the HVF-strong and HVF-weak subsamples are derived from the same distribution. Classifications made from the near-maximum-light spectra yield a similar result, which is consistent with previous studies \citep{mag12,ch14,mag14,pa15,sil15}. Note that the R$_{HVF}$-$\Delta$m$_{15}$(B) correlations shown for Si~II~6355 and Ca~II~IR triplet absorptions do not depend on the photospheric velocities.

\subsection{Peak B $-$ V color}\label{S_bv}
\begin{figure*}
\epsscale{.95} \plotone{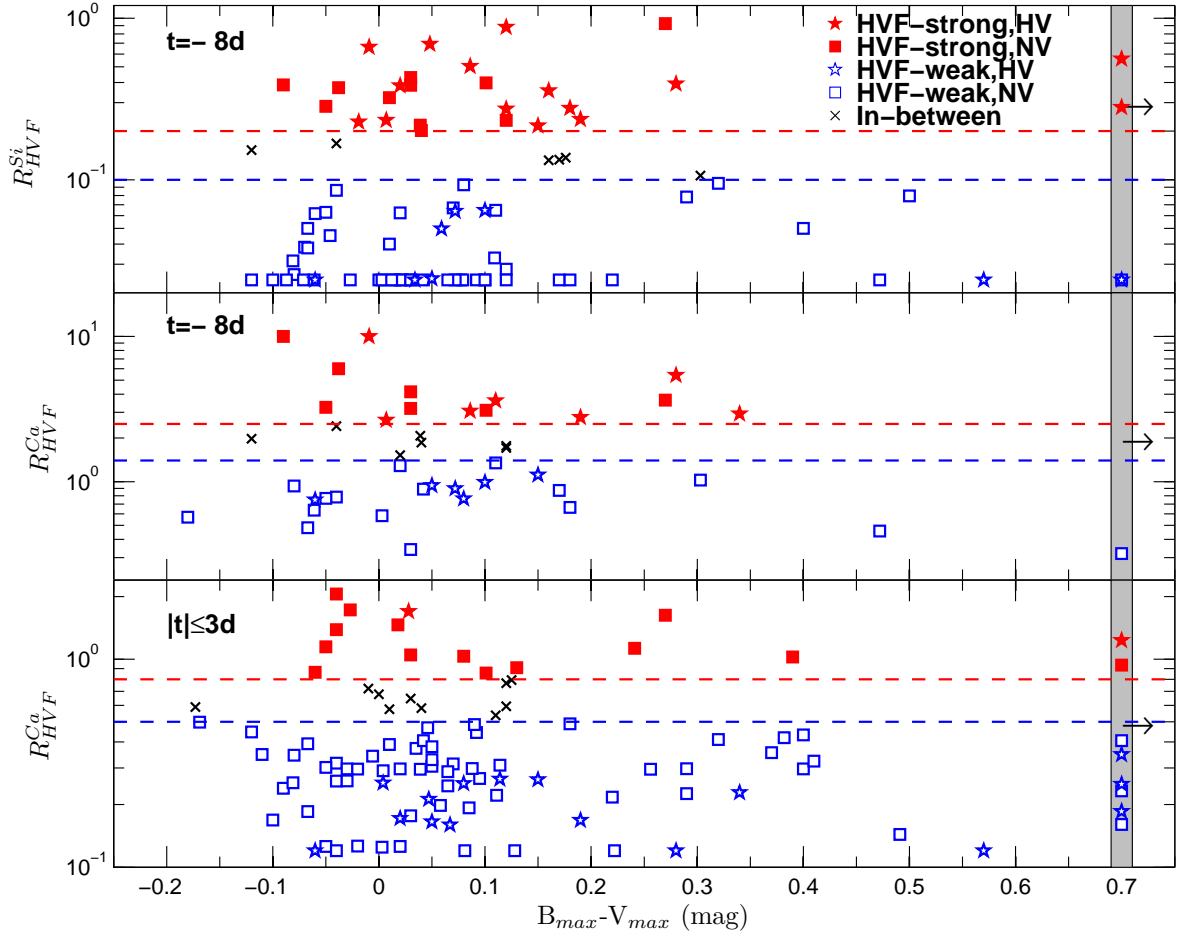}
\caption{\label{FigRhvf_bm-vm} The same plot of R$_{HVF}$ as Fig.\ref{FigRhvf_dm}, but against the B$_{max}$ $-$ V$_{max}$ color. For a better display, the SNe with B$_{max}$ $-$ V$_{max}\geq 0.4$ are shown in the yellow zone.}
\end{figure*}

\begin{figure*}
\epsscale{.95} \plotone{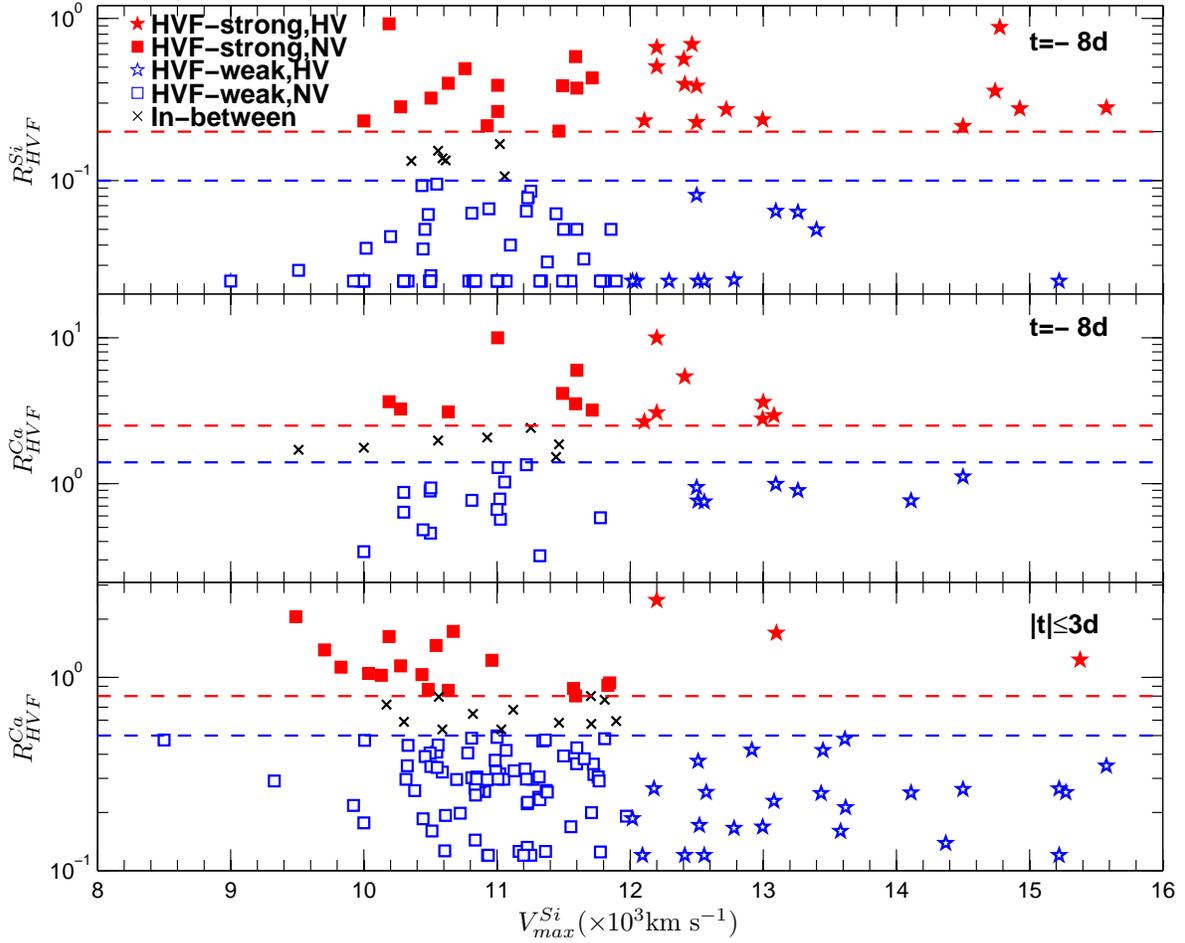}
\caption{\label{FigRhvf_vsi} The same plot of R$_{HVF}$ as Fig.\ref{FigRhvf_dm}, but against the Si~II velocities measured at around the maximum light.}
\end{figure*}

\begin{figure*}
\epsscale{.95} \plotone{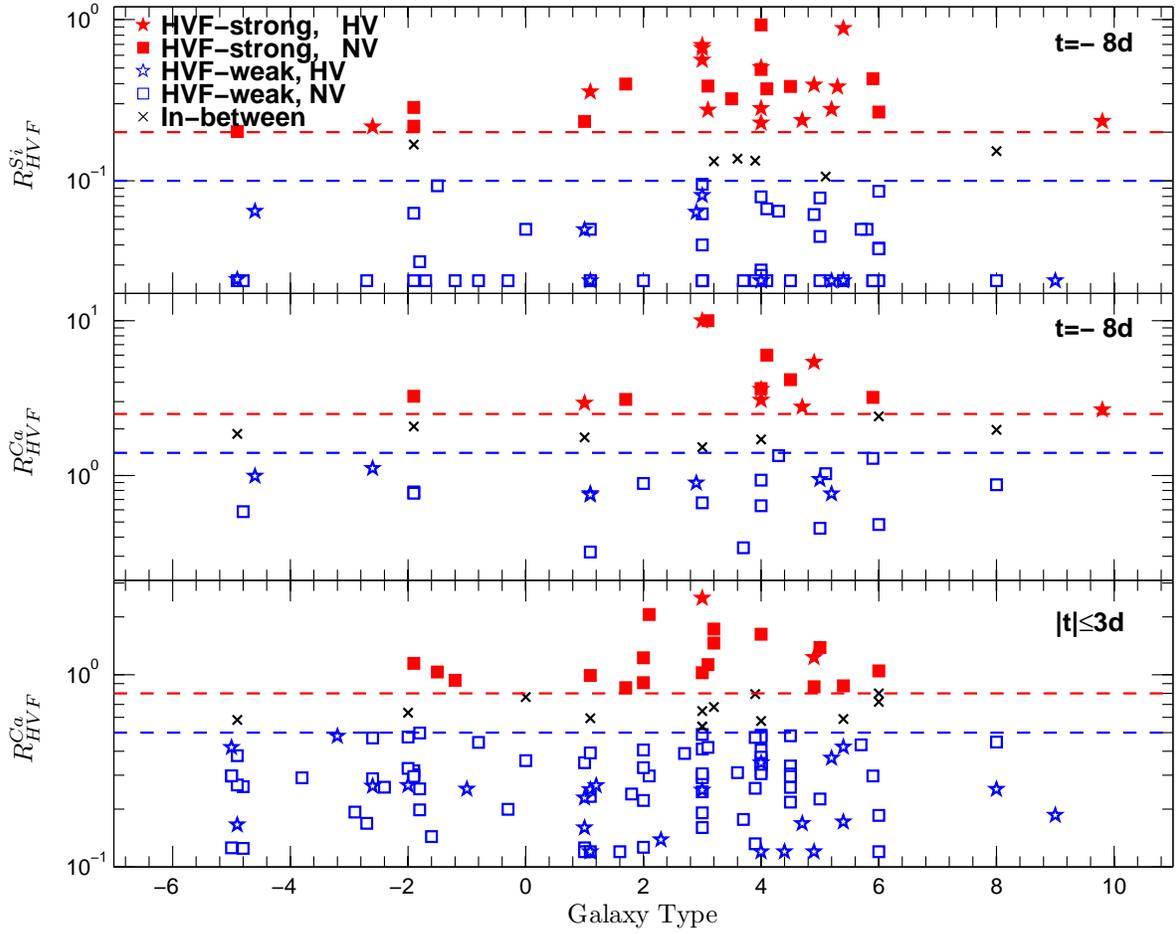}
\caption{\label{FigRhvf_Gtype}
The same plot of R$_{HVF}$ as Fig.\ref{FigRhvf_dm}, but against the morphology T types of the host galaxies.}
\end{figure*}

\begin{figure*}
\epsscale{.95} \plotone{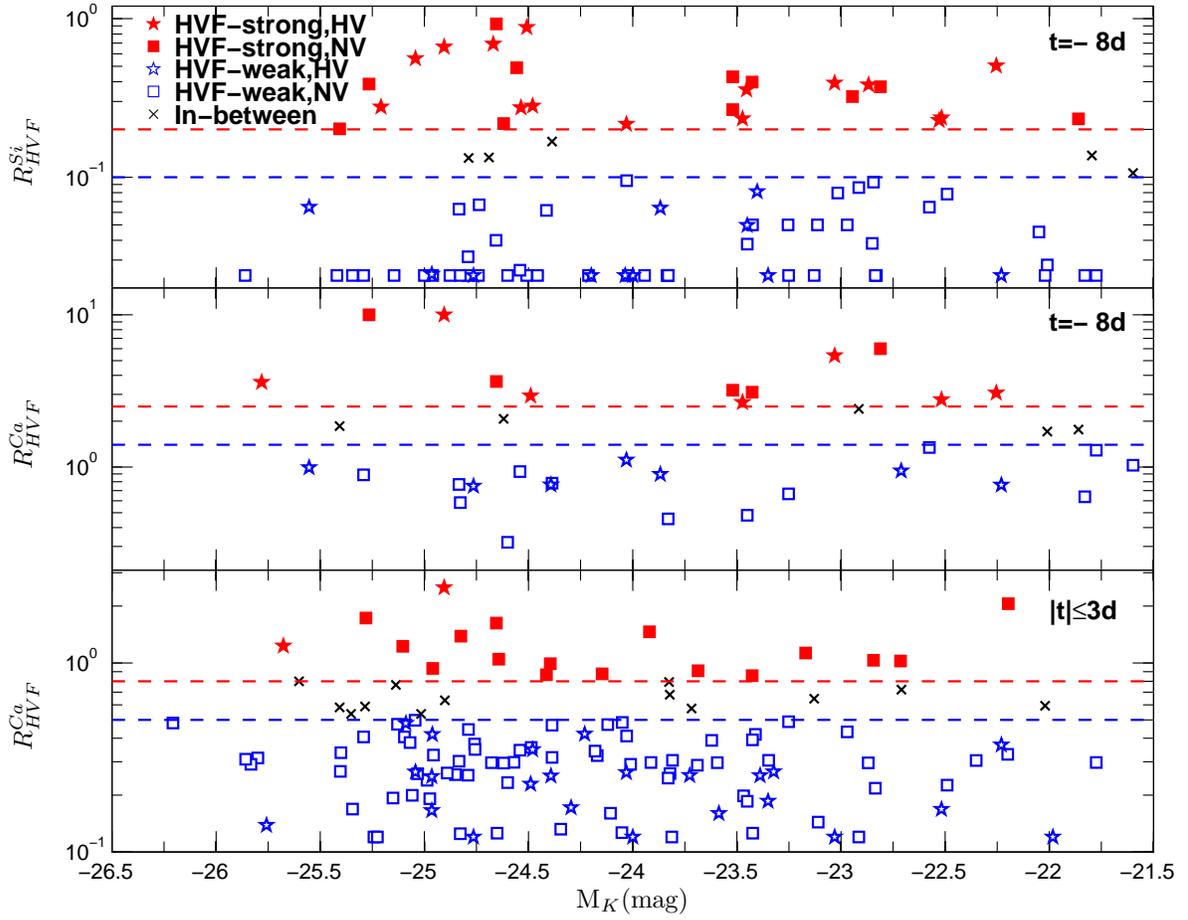}
\caption{\label{FigRhvf_Mk}
The same plot of R$_{HVF}$ as Fig.\ref{FigRhvf_dm}, but against the K-band absolute magnitudes of the host galaxies.}
\end{figure*}

\begin{figure*}
\epsscale{.95} \plotone{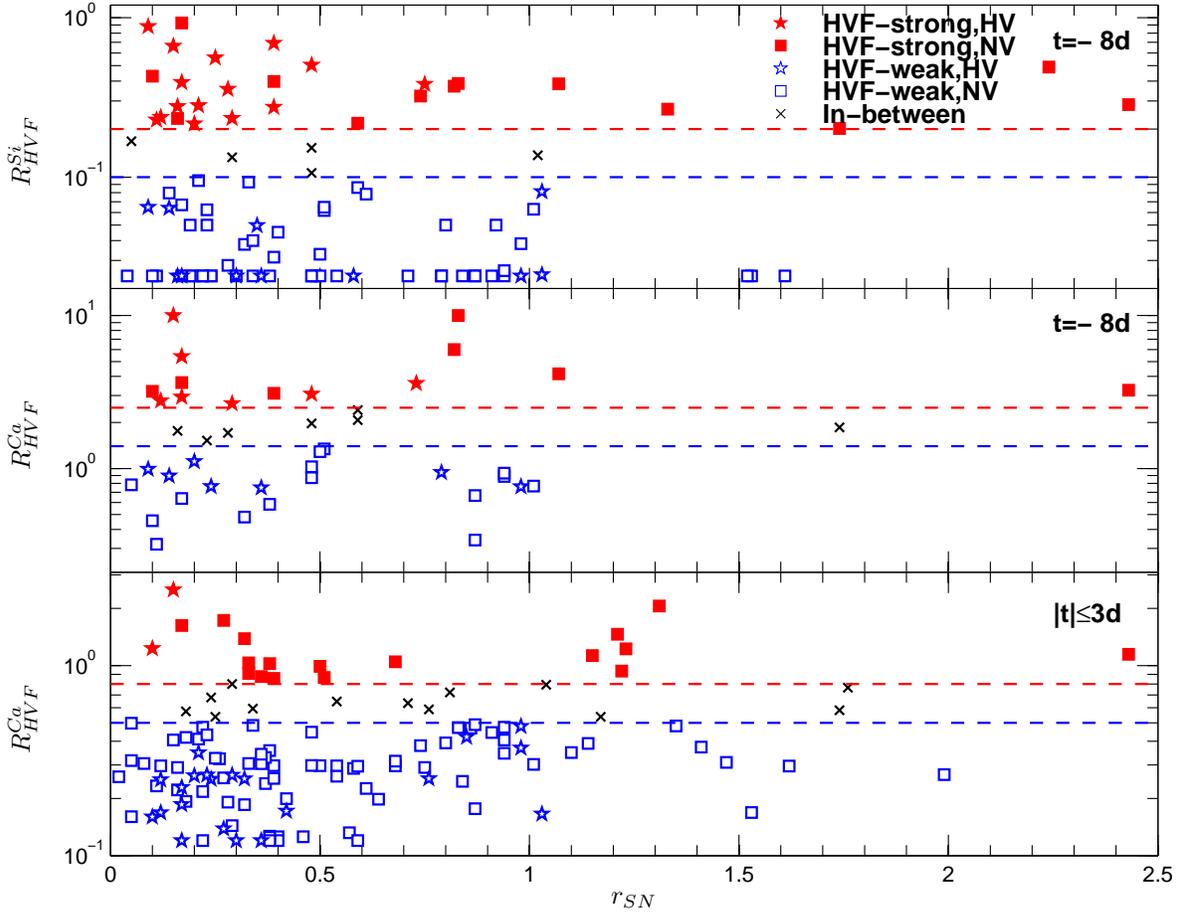}
\caption{\label{FigRhvf_rsn}
The same plot of R$_{HVF}$ as Fig.\ref{FigRhvf_dm}, but against the normalized radial distance of the SN from the center of the host
galaxies.}
\end{figure*}

\begin{figure*}
\epsscale{.95} \plotone{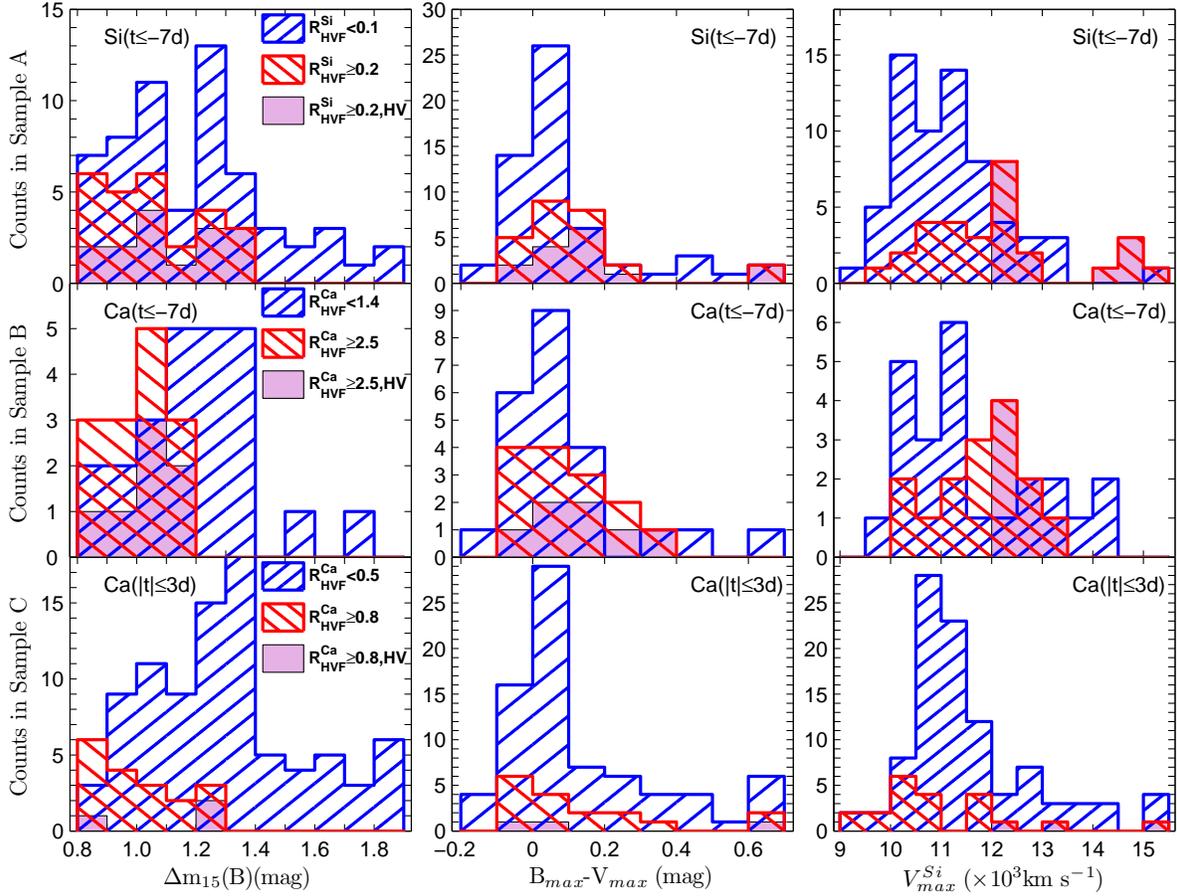} \caption{\label{FigHVF_ks}
Histogram distributions of some observables, including the light-curve decline rate ($\Delta$m$_{15}$(B)), the B $-$ V color
at the maximum light (B$_{max}$ $-$ V$_{max}$), the photospheric velocity at the maximum light $V^{Si}_{max}$. The sample is splitted                    into different subsamples based on the HVF parameter, R$_{HVF}$, and the photospheric velocity, $V^{Si}_{max}$. The upper panels show the                      distributions of the observables for the sample with measurements of the Si-HVFs from the early-time spectra (sample A);
the middle and lower panels show the cases for early-time Ca-HVFs (sample B) and maximum-light Ca-HVFs (sample C).}
 \end{figure*}

\begin{figure*}
\epsscale{.95} \plotone{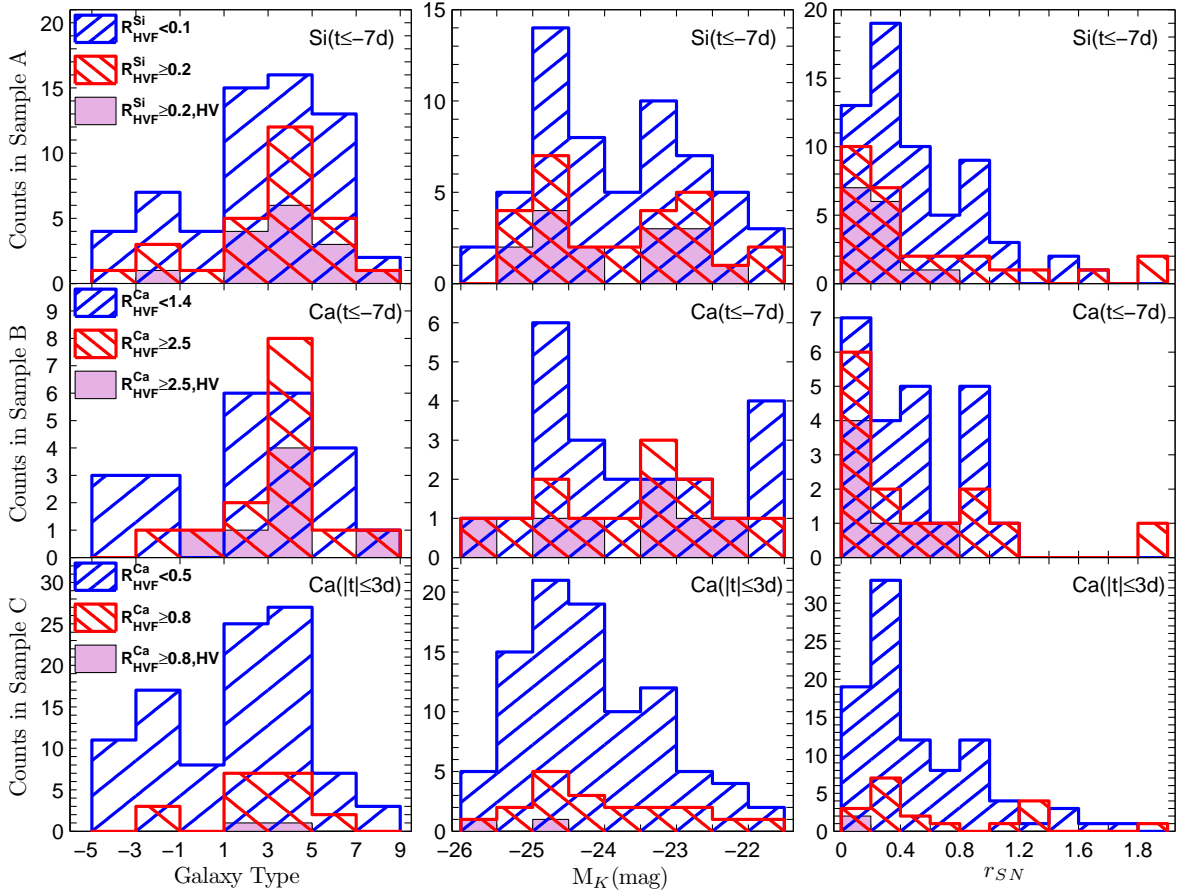}
\caption{\label{FigHVF_ks_gal} Histogram distributions of some parameters relevant with the host galaxies of SNe Ia, including the morphology T types, the K-band absolute magnitudes, and the normalized radial distance of the SN in its host galaxy (r$_{SN}$=R$_{SN}$/R$_{gal}$). The samples in different panels are defined in the same way as described in Fig.\ref{FigHVF_ks}.}
\end{figure*}

\begin{figure*}
\epsscale{.95} \plotone{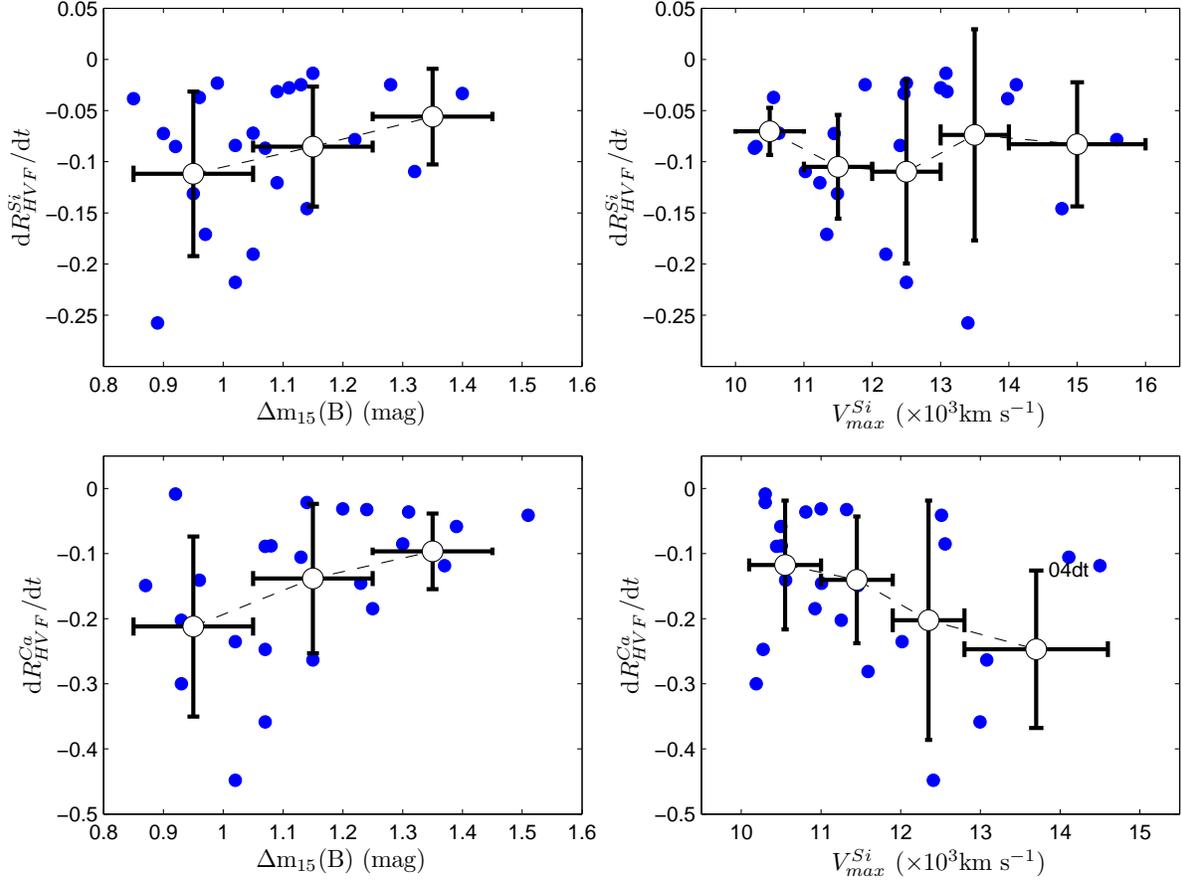}
\caption{\label{FigdRsidt} The decay rate of R$_{HVF}$ measured at t $<$ $-$ 7 days as a function of the light-curve decline rate $\Delta$m$_{15}$(B) (left panels) and maximum-light velocity $V^{Si}_{max}$ (right panels). For the measurements of dR$^{Ca}_{HVF}$/dt, the R$^{Ca}_{HVF}$ obtained at t $>$ $-$7 days are also included in the calculations but is represented with open symbols in the plot. The black circles show the decay rate averaged in binned $\Delta$m$_{15}$(B) space.}
\end{figure*}

The B $-$ V color is another important parameter that is related to the luminosity of SNe Ia \citep[i.e.,][]{tri98,wan05,gu07}. In Figure \ref{FigRhvf_bm-vm}, the HVF parameter R$_{HVF}$ is plotted as a function of the peak B $-$ V color (which was corrected for the Galactic reddening). One can see that neither the Si-HVFs nor the Ca-HVFs show significant correlation with the B $-$ V colors measured at around the maximum light, as also indicated by the larger P values of the K-S test. A similar conclusion has also been reached by \citet{ch14} using the Ca-HVFs detected in the near-maximum-light spectra.

Closer inspections of the distribution of Si-HVFs (Figure \ref{FigRhvf_bm-vm}), however, reveals that the HVF-strong SNe Ia with larger
photospheric velocity $V^{Si}_{max}$\footnote{Note that $V^{Si}_{max}$ is referred to the velocity of Si~II~6355 measured near the maximum light.}(HVF-strong HV SNe Ia) have apparently redder B$_{max}$ $-$ V$_{max}$ colors than the counterparts with lower velocities (HVF-strong NV SNe Ia), similar to that seen in the sample differentiated only by the photospheric velocity \citep{wa09a}. For example, the mean $B_{max} - V_{max}$ color is found to be 0.28$\pm$0.39 mag for the HV subgroup and 0.01$\pm$0.10 mag for the NV subgroup (see Figure \ref{FigHVF_ks}). This difference is larger than that obtained for the whole HV and NV SNe Ia of our sample (i.e., 0.23$\pm$0.33 mag versus 0.08$\pm$0.25 mag). These comparisons imply that the SNe Ia with both prominent Si-HVFs and higher $V^{Si}_{max}$ likely occurred in dusty environments, while those with prominent HVFs but lower $V^{Si}_{max}$ may come from relatively clean environments. Such a velocity-dependent color difference also exists among the sample of early-phase Ca-HVFs (sample B), whereas it is not known whether it holds for the sample from the maximum-light Ca-HVFs due to that there are few HV objects included in this sample.

\subsection{Photospheric Velocity $V^{Si}_{max}$}\label{S_vbmax}
Besides the light-curve decline rate and the peak $B_{max} - V_{max}$ color, the photospheric velocity and its gradient have been proposed as additional important parameters to describe the diverse properties of SN Ia explosion \citep[i.e.,][]{be05,wa09a}. Therefore we also examined the possible correlation between R$_{HVF}$ and the photospheric velocity $V^{Si}_{max}$, as shown in Figure \ref{FigRhvf_vsi}.

For the SN Ia sample defined by Si~II absorptions in the early phase, more than half of those with R$^{Si}_{HVF}$ $\gtrsim$ 0.2 at t$\sim-$8 days can be put into the HV group with v$_{Si}\gtrsim$12,000 km s$^{-1}$ at maximum light. A similar conclusion has been also obtained by \citep{sil15}. Owing to that only a smaller portion of the SN Ia sample have spectra covering the near-infrared wavelengths, it is not clear whether the sample defined by early-time Ca~II IR triplet maintains such a higher fraction of HV SNe Ia. Examining the near-maximum-light Ca~II~IR triplet, however, we found that the strong HVFs tend to occur in SNe Ia with smaller photospheric velocities (i.e., $V^{Si}_{max}$ $<$ 12,000 km s$^{-1}$). This is consistent with the results given by some recent studies \citep{ch14, pa15}. One reasonable explanation for the velocity difference between the HVF-strong SNe Ia selected from early-time Si~II and near-maximum-light Ca~II absorptions is that the HVFs in HV SNe Ia become weak more rapidly than those in NV SNe Ia (see Figure \ref{FigdRsidt} and discussions in \S3.6). In addition, line blending effect could be another possible reason. For the HV subclass, the HVFs can be easily blended with the photospheric components occurring at higher velocities.

Figure \ref{FigHVF_ks} compares the histogram distributions of these parameters of $\Delta$m$_{15}$(B), B$_{max}$ $-$ V$_{max}$, and $V^{Si}_{max}$ for SNe Ia  with and without HVFs. The detailed K-S test results for different samples and subsamples are listed in Table \ref{T_ks}.

\subsection{Explosion Environments}
We also examined the correlations of the HVFs of SNe Ia with properties of their host galaxies, including the morphology T-type and luminosity
of the host galaxy as well as the radial distance within the hosts, which are shown in Figure \ref{FigRhvf_Gtype}, Figure \ref{FigRhvf_Mk}, and Figure \ref{FigRhvf_rsn}, respectively. The luminosity of the host galaxies is represented by using their K-band absolute magnitudes from the 2MASS $Redshift$ $Survey$ \citep{hu12}, which has been corrected for the Galactic extinction. The radial distance is calculated as r$_{SN}$=R$_{SN}$/R$_{gal}$, where R$_{gal}$ represents the angular radius of the host galaxies and R$_{SN}$ is the observed angular distance of the SN from the galactic center.

In general, the HVF-strong and HVF-weak samples selected from Si~II or Ca~II lines do not show significant dependence on morphology or luminosity of the host galaxies, as suggested by the K-S test results (see also Table \ref{T_ks}). The dichotomy of the observed HVFs in SNe Ia does not seem to have a connection with their spatial positions either. Splitting the HVF-strong sample into the HV and NV subgroups does not reveal specific tendencies in distributions of the galaxy types or luminosity, but significant differences can be found in the radial distribution. Such discrepancies are clearly seen in the histogram distributions of different subsamples, as shown in Figure \ref{FigHVF_ks_gal}. The HVF-strong HV SNe Ia are found to have an averaged radial distance of r$_{SN}\sim$0.27$\pm$0.17, while the HVF-strong NV SNe Ia have a corresponding value of r$_{SN}$$\sim$1.04$\pm$0.76. Note that the mean r$_{SN}$ derived for all of our sample, the HV-weak subsample, and the HVF-strong subsample is comparable, which is 0.55$\pm$0.44, 0.53$\pm$0.37, and 0.59$\pm$0.64, respectively. This means that the HV SNe Ia of the HVF-strong sample prefer to occur in the inner regions of the host galaxy while the NV subsample show a contrary trend, which is consistent with that seen in the B$_{max}$ $-$ V$_{max}$ distribution. For these two subclasses from the sample with strong Si-HVFs, a K-S test gives a probability of 0.4\% that they come from the same population. For the case of early-phase Ca II, this probability is 11.1\%. These results imply that the HV and NV subclasses of the HVF-strong SNe Ia may have different progenitor populations. For example, the progenitor systems of the HV SNe Ia may have higher metallicity as suggested by \citet{wa13}.

\subsection{Temporal Evolution of the HVFs}\label{S_evolution}
Temporal evolution of the HVFs and its correlations with some observables (i.e., $\Delta$m$_{15}$(B) and $V^{Si}_{max}$) of SNe Ia provide additional information about the properties of the HVFs. Since our sample have a better early-phase spectroscopic coverage, we attempt to quantify the parameters such as the decay rate of R$_{HVF}$ and the epoch "when" R$_{HVF}$ drops below the cutoff values.

We first calculated the rates at which the HVFs decline in the early phases. The sample shown in Fig.\ref{FigdRsidt} contain those SNe Ia with $\gtrsim$ 3 spectra at t $<$ $-$7 days, and a linear function is assumed for the evolution of R$_{HVF}$ during the period at t $<$ $-$7 days. It can be found that the decay rate of the HVFs shows rough correlations with both the $\Delta$m$_{15}$(B) and the $V^{Si}_{max}$, with larger gradients (dR$_{HVF}$/dt) being measured for SNe Ia with smaller decline rates or larger velocities. Note that there is large scatter in these correlations, which can be in part attributed to that not all of the fast-expanding SNe Ia or slow decliners have strong HVFs and hence large dR$_{HVF}$/dt. On average, the Ca-HVFs tend to have larger decay rates than do the Si-HVFs for given $\Delta$m$_{15}$(B) or $V^{Si}_{max}$. These results complicate the observed effects of $\Delta$m$_{15}$(B) and $V^{Si}_{max}$ on the HVFs and its evolution.

\begin{figure*}
\epsscale{.95} \plotone{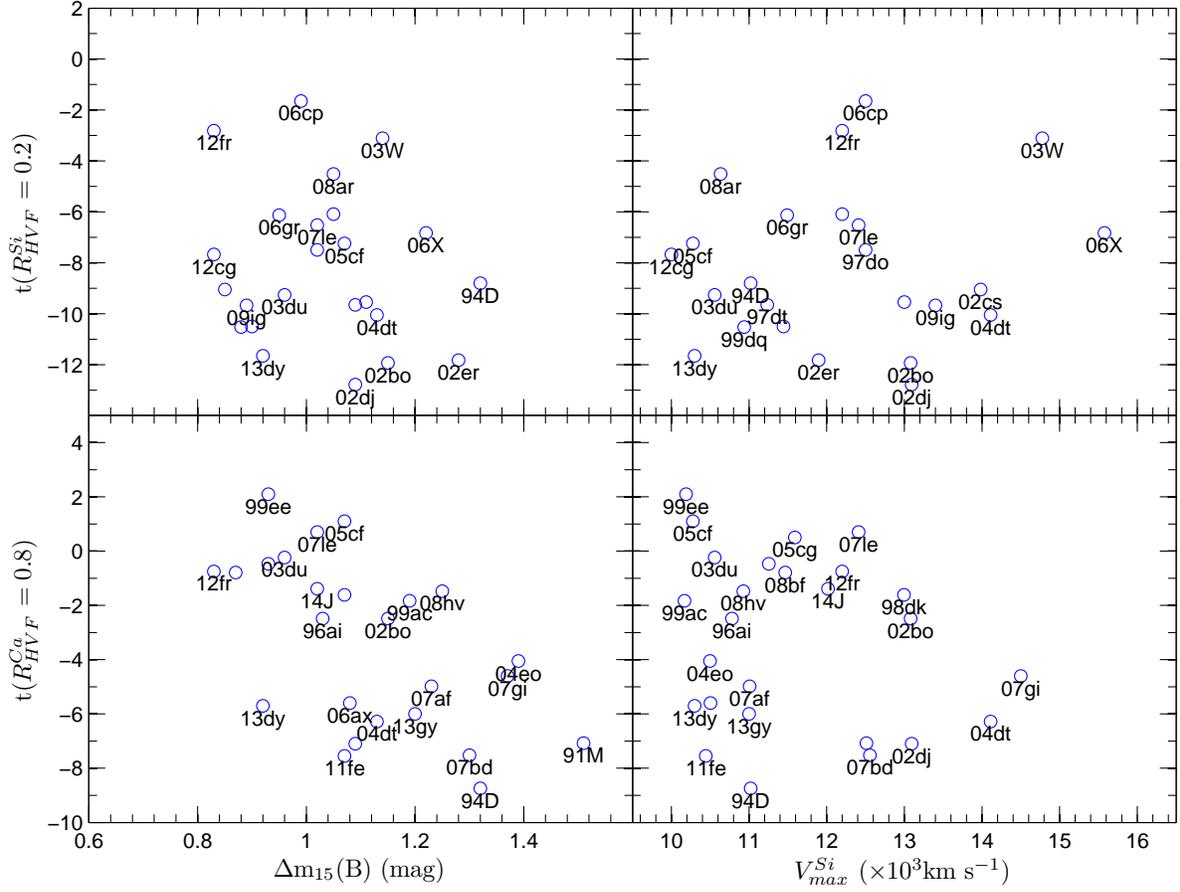} \caption{\label{Figtcut_si} The distribution of the epoch determined for a given SN when its HVFs become "weak" (i.e., R$^{Si}_{HVF}$ $<$ 0.2 or R$^{Ca}_{HVF}$ $<$ 0.8) is plotted versus $\Delta$m$_{15}$(B) and $V^{Si}_{max}$. The decay rates obtained in Fig.\ref{FigdRsidt} are used for extrapolation of the cutoff time when necessary.}
\end{figure*}

With the decay rates of the HVFs, we can then determine the cut-off time (t$_{cut}$) when the R$_{HVF}$ becomes "weak" for a given SN. As shown in Fig.\ref{Figtcut_si}, this epoch estimated for R$^{Si}_{HVF}$ = 0.2 could vary from $-$13.0 days to $-$2.0 days for different SNe Ia, and the corresponding epoch derived for R$^{Ca}_{HVF}$ = 0.8 has a distribution ranging from $-$9.0 days to +2.0 days relative to the maximum light. Despite of the large scatter, there are correlations between the cut-off time of these two criteria and the decline rate $\Delta$m$_{15}$(B). The SNe Ia with larger $\Delta$m$_{15}$(B) tend to reach at their t$_{cut}$ time earlier than those with smaller $\Delta$m$_{15}$(B). In contrast, no significant correlation is found between t$_{cut}$ and $V^{Si}_{max}$.

\subsection{Correlation of the HVFs of Si~II$\lambda$6355 and Ca~II~IR Triplet}
Both previous studies and our analysis show that the HVFs are more commonly detected in the Ca~II~IR triplet than in the Si~II line for most SNe Ia. Among our sample with t$<$$-$7 day spectra, almost all SNe Ia (except for SN 2002cs) are found to have R$^{Ca}_{HVF}$ $>$ 0.2 in the early phase, while this fraction is only 27.1\% (29 out of 107) for the Si~II line. It's thus interesting to examine whether the Si~II and Ca~II absorptions are correlated in velocity and strength.

\begin{figure*}
\epsscale{.95} \plotone{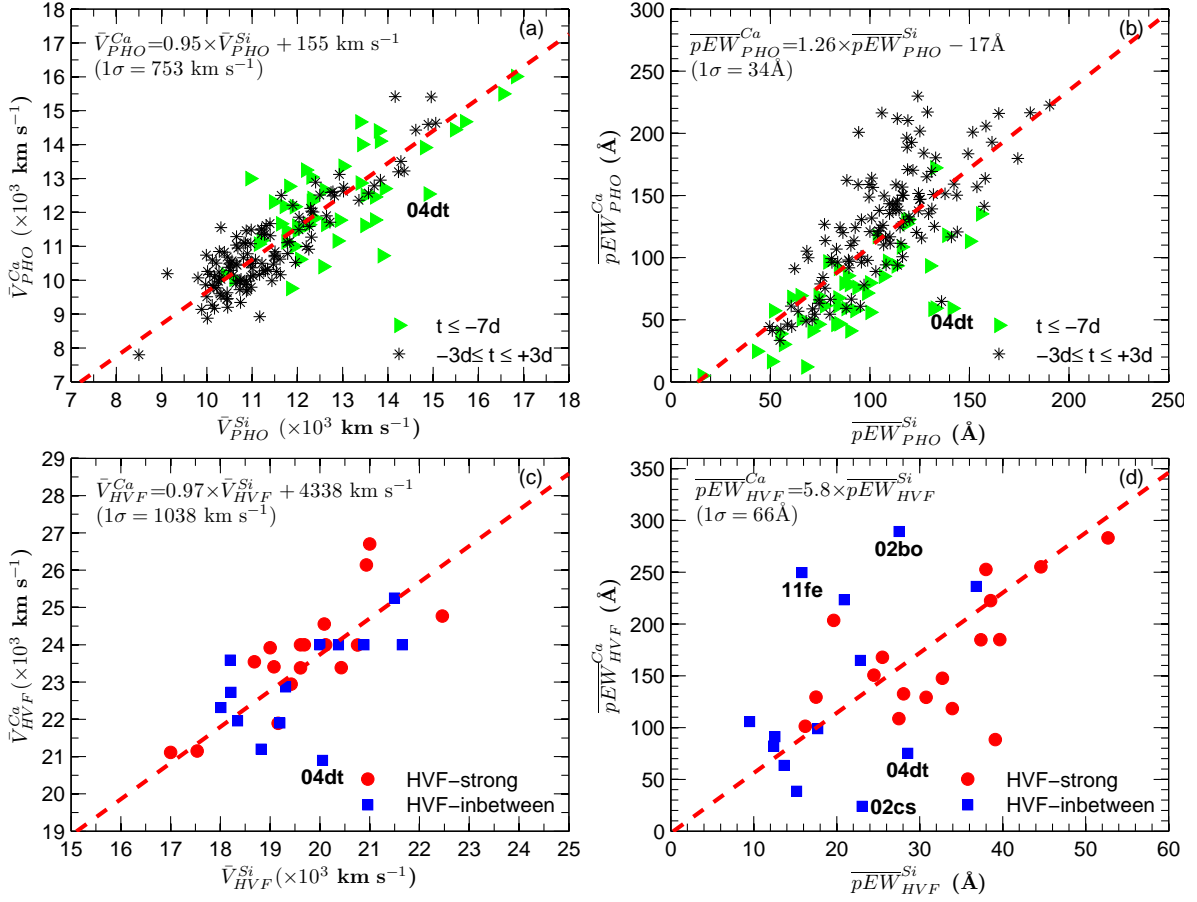} \caption{\label{FigCaSi} Correlations of the velocity and absorption strength (pEW) between Si~II 6355 line and Ca~II~IR triplet. The upper panels (a and b) show the photospheric components measured at around the maximum light, the lower panels (c and d) shows the correlations of the HVF components measured at t $< -$7 days. The dashed line represents the best linear fit to the data, with function forms shown in each panel. The solid lines are drawn to guide the eyes, with scales of 1:1 for panels a and b, 1:1 for panel c, and 1:6 for panel d, respectively.}
\end{figure*}

Figure \ref{FigCaSi}a and b show the comparison of the photospheric components of Si~II~6355 and Ca~II~IR triplet. One can see that these two lines have similar photospheric velocities in both early and near-maximum-light phases, and they also show a correlation in the absorption strength (e.g., pEW). Note that the photospheric component of Si~II~6355 absorption is on average stronger than that of the Ca~II~IR triplet at t $<$ $-$7 days and it tends to become relatively weak at around the maximum light. We compare the HVFs of Si~II and Ca~II in Figure \ref{FigCaSi}c and d. At earlier phases, the Ca-HVFs have velocities that are apparently higher than the Si-HVFs (e.g., $V_{HVF}$(Ca) $\approx$ $V_{HVF}$(Si) + 4300 km s$^{-1}$); and they also appear much stronger at similar phases, with the pEW of the Ca-HVFs being roughly about six times that measured for the Si-HVFs.  These results indicate that the Ca-HVFs form at a velocity of about a few thousand km s$^{-1}$ above the Si~II layer, and its formation seems to be easier than the Si-HVFs. Although some correlations can be seen in the HVFs of Si and Ca, they show larger scatter (especially in the absorption strength) compared to that seen in the photospheric components.

\begin{figure*}
\epsscale{.95}
\plotone{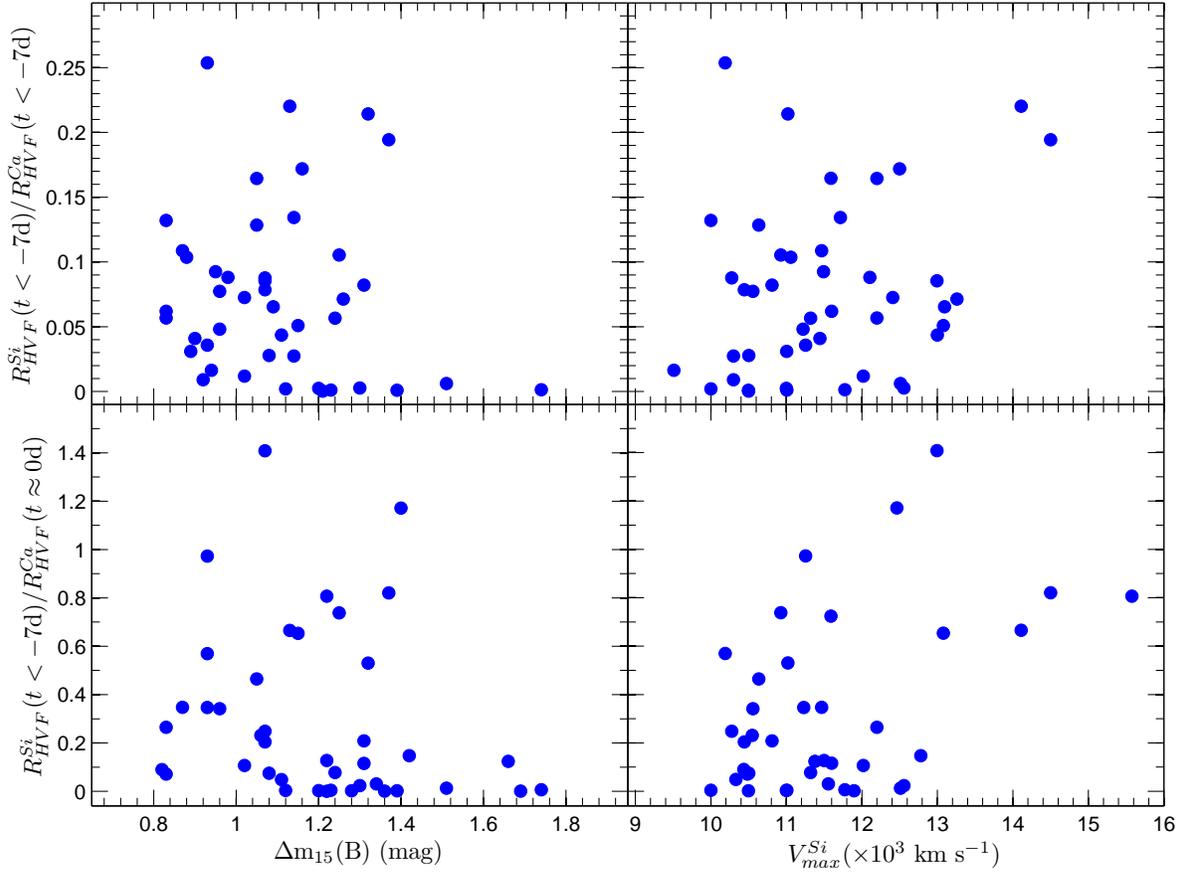}
\caption{\label{FigRsiRca} The ratio of R$^{Si}_{HVF}$ and R$^{Ca}_{HVF}$ is plotted versus $\Delta$m$_{15}$(B) and $V^{Si}_{max}$. Upper panels: the measurements obtained at early phases (i.e., t$\sim$ $-$ 8 days). Lower panels: the measurements made between R$^{Si}_{HVF}$ at early phases and R$^{Ca}_{HVF}$ at maximum light.}
\end{figure*}

To further examine the scatter in absorption strength of Si- and Ca-HVFs, we perform some additional tests by plotting the ratio of their relative strengths as a function of $\Delta$m$_{15}$(B) and $V^{Si}_{max}$. The upper panels of Figure \ref{FigRsiRca} show R$^{Si}_{HVF}$ at early phases divided by R$^{Ca}_{HVF}$ at early phases (i.e., t$\sim$ $-$ 8 days), and the lower panels of the plot show the distribution of R$^{Si}_{HVF}$ at early phases divided by R$^{Ca}_{HVF}$ at maximum light. One prominent feature of these plots is that the R$^{Si}_{HVF}$/R$^{Ca}_{HVF}$ ratio tends to be smaller for SNe Ia with larger decline rates (i.e., $\Delta$m$_{15}$(B) $>$ 1.40 mag) while this ratio is found not to be the largest for SNe Ia
with lowest $\Delta$m$_{15}$(B). This result indicates that the temperature of the HVF zones should be in a reasonable range, and too-high or too-low temperature would make it difficult to form prominent Si-HVFs. On the other hand, the R$^{Si}_{HVF}$/R$^{Ca}_{HVF}$ ratio obtained at early phases does not seem to depend on the photospheric velocity but a possible correlation may exist for the ratio of early-phase R$^{Si}_{HVF}$ to maximum-light R$^{Ca}_{HVF}$, with fast-expanding SNe Ia having on average relatively stronger Si-HVFs. This may be explained with the fact that the Ca-HVFs of HV SNe Ia decline at a faster pace than the corresponding Si-HVFs (see \S\ref{S_evolution}). At given $\Delta$m$_{15}$(B) or $V^{Si}_{max}$, the observed diversity of the HVFs can be related to the differences in the explosion physics and/or progenitor stars (see discussions in \S \ref{S_origin}).

\section{Origins of the HVFs in SNe Ia}\label{S_origin}
Results of spectral measurements presented in this paper provide some clues to the origins of the HVFs of Si II and Ca II seen in some SNe Ia. There are a few popular explanations for the formation of the HVFs in SNe Ia -- the abundance enhancement (AE) scenario \citep{ma05a,ma05b,ro06,mae10,ch13,ch14,bl13,no13,mag14}, the density enhancement (DE) scenario
\citep{ge04,ma05a,ma05b,ta06,al07,ta08,to12,ch13,ch14,mag14}, and the ionization effect (IE) scenario \citep{ma05a,ta08}.

In the delayed-detonation scenario \citep{kh91}, a substantial mixing is expected to take place in the first deflagration phase,
thus leading to an asymmetry in the distribution of the detonation ignition. Therefore, the detonation could burn C/O into Si/Ca in
some directions even in the outermost layers \citep{mae10, sei13}. The similar situation may also be realized in a violent merger of
two WDs, where the `strong' detonation side will produce Si/Ca more effectively than the other side \citep{ro12}. The origin of HVFs might not necessarily involve an asymmetric explosions as in these examples, and such an example includes a double-detonation model \citep[e.g., ][]{nom80,woo80,nom82a,fin07,woo11}. This is a model where the explosion is initiated by the He-burning near the WD surface, thus producing
abundance enhancement in this region which might be observed as the HVFs \citep[see e.g., ][]{mag14}.

In the DE scenario, the HVFs are suggested to originate from a high-density shell at large radii while the abundance there is typical of the C/O-rich layer of the expanding ejecta. This shell may be formed from interaction between the SN ejecta and CSM \citep{ge04,ma05b,mul15,ta06,ta08}. This may happen if the exploding WD is surrounded by relatively dense CSM at the vicinity as related to the pre-explosion WD activity or even to an explosion process \citep{hof95,hof96}.

In the IE scenario, a small amount of H in the outermost layer serve as a source of free electrons, and thus suppressing the ionization status of Ca and Si through recombination. This then leads to a larger amount of Ca~II and Si~II, potentially producing the HVFs \citep{ma05a,ta08}. This may happen either as a contamination of H in the WD surface before the explosion or due to an interaction between the ejecta and the H-rich CSM, which is similar to the DE scenario.

The information on velocity structures of different elements should in principle help to distinguish the above scenarios. The observational finding that the HVFs of Si~II and Ca~II lines show correlations in both velocity and strength would indicate that the these HVFs may be formed in the same layer, and the relative abundance ratio between Si and Ca in the outermost HVF-forming layer has a generic value for different SNe Ia. The observation results that the HVFs are more commonly detected in Ca~II than in Si~II line (i.e., generally a larger pEW in the Ca-HVFs than Si-HVFs), as well as that the Ca-HVFs which formed at higher velocity than the Si-HVFs, might well be explained primarily by the different excitation energies required to produce lines of Si~II~6355 ($E_{\rm ex} = 8.12$ ev) and Ca~II~IR triplet ($E_{\rm ex} = 1.7$ ev). For relatively low temperature expected for the outermost HVF-forming layer, Ca~II~IR absorption is more easily formed and saturated, and this tendency is expected to be strengthened toward the higher velocity.

The association of strong Si-HVFs with HV SNe Ia and the opposite tendency in maximum-light Ca-HVFs could be understood by an ionization effect ($E_{\rm ion} = 16.3$ ev for Si and $E_{\rm ion} = 11.9$ ev for Ca, for the ionization from a singly ionized ion to a doubly ionized). A fraction of Si~II is sensitively dependent on the radiation temperature and decreases quickly for higher temperature \citep{hach08}, while a fraction of Ca~II can be fairly constant for a wide range of temperature \citep{ta08}. Therefore, Si~II~6355 could become weaker for higher temperature despite its large excitation energy, while Ca~II~IR would not be sensitively dependent on the temperature (also due to low $E_{\rm ex}$ which saturates the line relatively easily). Given a similar luminosity for the NV- and HV-SNe, the former has a higher radiation temperature due to slower
expansion, and it is thus more difficult to produce the Si~II line at higher velocities.

On the other hand, the trend that the HVFs are more strongly tied to SNe Ia with smaller $\Delta m_{15}$(B) is difficult to be interpreted as the ionization and/or excitation effects alone. These slow decliners usually have higher luminosities and photospheric temperatures, and therefore may well suppress the formations of HVFs through the ionization and excitation effects. Therefore, the difference in the strength of the HVFs here should reflect the difference in the structure of the HVF-forming region. Namely, these HVF-strong SNe should have more `burned material' in the outer most layer in the AE scenario, or more substantial `dense shell' in the DE scenario.

Moreover, the observation result that the HVF-strong HV SNe Ia tend to occur in the inner regions of the host galaxies and have redder B$_{max}$ $-$ V$_{max}$ colors and the opposite trend seen in the HVF-strong NV objects perhaps indicate a physical difference in the outermost layer in these two subgroups showing HVFs, beyond what only the thermal effects for the same configuration could explain. The redder color of the HV SNe Ia could be interpreted as a larger line-of-sight reddening or higher metallicity of the progenitor stars, consistent with that they may arise from younger stellar populations \citep[i.e., ][]{wa13}. This link is further supported by a possible association between SNe with small $\Delta$m$_{15}$(B) (thus bright) and the presence of Na absorption presumably from CSM \citep{mag14}. Thus it is possible that the HVF-strong HV SNe are surrounded by a denser CSM, resulting in a more effective creation of a dense shell, along with the DE scenario. Whereas the HVF-strong NV SNe may burn their outer layers more effectively than the other population, along with the AE scenario. This link could also be explained by the IE scenario; for example, a small amount of H in the environment around the young population may contaminate the outer layer of the ejecta, thus making the HVFs along the line of the IE scenario.

While the observational information and our understanding of the formation of the HVFs are still limited to distinguish different scenarios, our analysis adds some insights. Regarding the AE scenario, according to the multi-dimensional delayed detonation models \citep{mae10,sei13}, more asymmetric configuration in the detonation trigger, following deflagration, leads to brighter SNe and stronger burning of the outermost layer. This expectation is consistent with the observed trend for normal SNe Ia, but the DE scenario seems to be more favorable to the HV SNe Ia.

\section{Conclusions}\label{S_conclusion}
A large sample of SNe Ia including very early-time spectra were collected to examine the HVFs in Si~II~6355 line and Ca II NIR triplet. Multiple-component Gaussian fits were applied to detect such features, with better determinations of the pseudo continuum around the absorption features and photospheric velocities from the HVF-free absorption features such as Si~II~5972. Our analyses show that the photospheric components of Si~II~6355 and Ca~II~IR triplet absorptions are well correlated in velocity and strength. A linear correlation is found to also exist in the velocity of their HVFs, but the Ca II velocity is systematically higher than the Si II by about 4,000 km s$^{-1}$. Moreover, the strength of the HVFs of Si II and Ca II shows larger scatter in comparison with that measured for the photospheric component. We speculate that these features and correlations are explained by different ionization and thermal processes. The HVF-forming regions are seemingly shared by the Si~II and Ca~II, and the relative abundance between Si and Ca seems to have a universal value.

Based on the ratio of the absorption strength between HVFs and the photospheric components (e.g., defined as R$_{HVF}$), we divided the SN Ia sample into the HVF-strong and the HVF-weak subclasses according to the early-time Si~II~6355 and Ca~II~IR triplet absorptions as well as
the maximum-light Ca~II~IR triplet. We examined correlations of properties of the HVFs with other observables. The HVF-strong SNe Ia are
found to have smaller light-curve decline rates (e.g., $\Delta$m$_{15}$(B) $\lesssim$ 1.4), while the HVF-weak group seem to have a broader
distribution of $\Delta$m$_{15}$(B), consistent with previous results \citep{mag12, ch14, mag14, sil15}. On the other hand, we found that the HV-SNe Ia with prominent Si-HVFs at early phases have redder peak B $-$ V colors and tend to occur near the galactic center, while the HVF-strong NV group show an opposite trend. These correlations we found in the statistical analysis indicate that there are real differences in the HVF forming region beyond the ionization/thermal effects as we discussed in section \S\ref{S_origin}.

Even adding the results in present paper, the origin of the HVFs is not yet clarified. The observations are consistent with two scenarios: (1) the HVF-forming region is created by the explosion mechanism, especially by the burning induced by the detonation wave, and (2) it is a dense shell created by the interaction between the SN ejecta and CSM.

\acknowledgments We thank the anonymous referee for his/her insightful suggestions which help improve the paper a lot. The work is supported by
the Major State Basic Research Development Program (2013CB834903), the National Natural Science Foundation of China (NSFC grants 11178003 and
11325313), the Foundation of Tsinghua University (2011Z02170), China Scholarship Council (CSC 201406210312), and the Strategic Priority Research Program "The Emergence of Cosmological Structures" (Grant No. XDB09000000) of the Chinese Academy of Sciences. The work by K.M. is partly supported by JSPS Grant-in-Aid for Scientific Research ns (No. 26800100) and by World Premier International Research Center Initiative (WPI Initiative), MEXT, Japan. Funding for the LJ 2.4-m telescope has been provided by CAS and the People's Government of Yunnan Province. This research has made use of the CFA Supernova Archive, which is funded in part by the US National Science Foundation through grant AST 0907903. This research has also made use of the Lick Supernova Archive, which is funded in part by the US National Science Foundation.

\cleardoublepage

\clearpage

\begin{deluxetable}{lccccccccccccccccccc}
\centering\tabletypesize{\scriptsize}\setlength{\tabcolsep}{0.02in}  
\tablecaption{Photometric and Spectroscopic Parameters of Type Ia
Supernovae \label{T_par}} \tablewidth{0pt} \tablehead{
 \colhead{SN}
&\colhead{Subclass\tablenotemark{a}}
&\colhead{$\Delta$m$_{15}$(B)\tablenotemark{b}} &\colhead{B$_{max}$$-$V$_{max}$\tablenotemark{c}} &\colhead{$V_{CMB}$\tablenotemark{d}}
&\colhead{Galaxy}\tablenotemark{e}
&\colhead{$r_{SN}$\tablenotemark{f}}
&\colhead{M$_{K}$(gal)\tablenotemark{g}}
&\colhead{HVF\tablenotemark{h}}
&\colhead{R$^{Si}_{HVF}$\tablenotemark{i}}
&\colhead{R$^{Ca}_{HVF}$\tablenotemark{j}}
&\colhead{R$^{Ca}_{HVF}$\tablenotemark{k}}
&\colhead{Ref.\tablenotemark{l}}
\\
 \colhead{} &\colhead{} &\colhead{(mag)} &\colhead{(mag)} &\colhead{(km s$^{-1}$)} &\colhead{type} &\colhead{} &\colhead{(mag)} &\colhead{type}
&\colhead{($-$8d)} &\colhead{($-$8d)} &\colhead{(0d)} &\colhead{} }

\startdata
1984A & HV & 1.22(10) & 0.16(08) & 64 & Sa(1.1) & 0.28 & -23.45(44) & Strong & 0.36 & \nodata & \nodata & 1 \\
1989B & N & 1.06(13) & 0.32(10) & 1075 & Sb(3.0) & 0.21 & -24.03(45) & Weak & 0.09 & \nodata & 0.41 & A;B;2 \\
1989M & N & 1.10(20) & \nodata & 1843 & Sb(3.0) & 0.28 & -24.97(31) & \nodata & \nodata & \nodata & 0.19 & B;3 \\
1990N & N & 0.95(05) & 0.01(06) & 1338 & Sbc(3.5) & 0.74 & -22.95(26) & Strong & 0.25 & \nodata & \nodata & A;B;4 \\
1991M & HV & 1.51(10) & \nodata & 2266 & Sc(5.2) & 0.98 & -22.23(51) & Weak & 0.00 & 1.26 & 0.17 & A;B;5 \\
1991T & 91T & 0.94(03) & 0.17(03) & 2079 & Sb(4.0) & 0.28 & -22.01(13) & Weak & 0.03 & 1.71 & \nodata & A;B;4 \\
1994D & N & 1.32(05) & -0.08(04) & 953 & S0(-1.9) & 0.05 & -24.39(41) & IB & 0.17 & 0.78 & 0.32 & A;B;6 \\
1994S & N & 0.97(04) & -0.03(06) & 4831 & Sab(2.0) & 0.38 & -24.05(14) & \nodata & \nodata & \nodata & 0.13 & A;B;7 \\
\enddata
\end{deluxetable}

\begin{table}
   \centering\caption[]{\centering{Selection and Subclassification Criteria of Different Samples}}
         \label{T_criterion}
     $$
         \begin{array}{p{0.2\linewidth}cccc}
            \hline
            \noalign{\smallskip}
    \tabincell{c}{Sample}                   & \tabincell{c}{Number} & \tabincell{c}{HVF-strong} & \tabincell{c}{HVF-weak} & \tabincell{c}{In-between}
    \\\hline
    \tabincell{c}{A(Si~II, $t<$-7d)}        & 107       & R^{Si}_{HVF}\geq 0.2  & R^{Si}_{HVF} < 0.1  & 0.1 \leq R^{Si}_{HVF} < 0.2
    \\
    \tabincell{c}{B(Ca~II, $t<$-7d)}    & 46        & R^{Ca}_{HVF}\geq 2.5  & R^{Ca}_{HVF} < 1.4  & 1.4 \leq R^{Ca}_{HVF} < 2.5
    \\
    \tabincell{c}{C(Ca~II, $|t|<$3d)}       & 138       & R^{Ca}_{HVF}\geq 0.8  & R^{Ca}_{HVF} < 0.5  & 0.5 \leq R^{Ca}_{HVF} < 0.8
    \\
            \noalign{\smallskip}
            \hline
         \end{array}
     $$
\end{table}

\begin{deluxetable}{lccccccccccccccccccc}
\tabletypesize{\scriptsize}\setlength{\tabcolsep}{0.02in}
\tablecaption{\label{T_sie}Fit results of the Si~II Absorption Features $\lambda6355$ and $\lambda5972$ from Sample A}
\tablewidth{0pt} \tablehead{\colhead{SN} &\colhead{Phase\tablenotemark{a}} &\multicolumn{5}{c}{Si~II$\lambda$6355} &\colhead{} &\multicolumn{2}{c}{Si~II$\lambda$5972}  \\
\cline{3-7} \cline{9-10} \\
\colhead{} &\colhead{} &\colhead{$V^{Si}_{PHO}$\tablenotemark{b}} &\colhead{$pEW^{Si}_{PHO}$\tablenotemark{c}} &\colhead{$V^{Si}_{HVF}$\tablenotemark{d}}
&\colhead{$pEW^{Si}_{HVF}$\tablenotemark{e}} &\colhead{$R_{HVF}$\tablenotemark{f}} &\colhead{} &\colhead{$V^{Si}_{HVF}$\tablenotemark{g}}
&\colhead{$pEW^{Si}_{PHO}$\tablenotemark{h}}

\\
 \colhead{} &\colhead{(days)} &\colhead{(km s$^{-1}$)} &\colhead{(\AA)} &\colhead{(km s$^{-1}$)} &\colhead{(\AA)} &\colhead{} &\colhead{} &\colhead{(km s$^{-1}$)} &\colhead{(\AA)}}

\startdata
1984A & -7.0 & 15394 & 139 & 21240 & 45 & 0.33 & ~ & 14358 & 39 \\
1989B & -7.0 & 11673 & 127 & 21000 & 8 & 0.07 & ~ & 10394 & 30 \\
1990N & -13.9 & 14634 & 62 & 21547 & 38 & 0.62 & ~ & 12633 & 5 \\
1990N & -6.9 & 11250 & 58 & 18694 & 10 & 0.17 & ~ & 11131 & 9 \\
1991M & -7.8 & 13384 & 118 & 18000 & 1 & 0.00 & ~ & 12484 & 29 \\
1991T & -11.5 & 15000 & 21 & 19422 & 0 & 0.00 & ~ & 9500 & 13 \\
1991T & -10.5 & 13247 & 15 & 17000 & 0 & 0.00 & ~ & 13317 & 10 \\
1991T & -9.5 & 10545 & 19 & 19000 & 1 & 0.03 & ~ & 13344 & 7 \\
\enddata

\tablecomments{}
\tablenotetext{a}{Days since the B-band maximum light;}
\tablenotetext{b}{Photospheric velocity of Si~II~6355 absorption;}
\tablenotetext{c}{Equivalent width of the photospheric (PHO) component of Si~II~6355 absorption;}
\tablenotetext{d}{Velocity of the high-velocity feature (HVF) of Si~II~6355 absorption;}
\tablenotetext{e}{Equivalent width of the high-velocity feature (HVF) of Si~II~6355;}
\tablenotetext{f}{R$_{HVF}^{Si}$ is defined as the ratio of the pEW of the HVF component to that of the PHO component;}
\tablenotetext{g}{Photospheric velocity of Si~II~5972, PHO component;}
\tablenotetext{h}{Equivalent width of photospheric component of Si~II~5972, PHO component.}

\end{deluxetable}

\clearpage

\begin{deluxetable}{lcccccc}
\tabletypesize{\scriptsize}
\tablecaption{\label{T_cae}Fit results of the Ca~II~IR triplet of Sample B} \tablewidth{0pt} \tablehead{
 \colhead{SN} &\colhead{Phase\tablenotemark{a}}
&\colhead{$V^{Ca}_{PHO}$\tablenotemark{b}} &\colhead{$pEW^{Ca}_{PHO}$\tablenotemark{c}} &\colhead{$V^{Ca}_{HVF}$}\tablenotemark{d}
&\colhead{$pEW^{Ca}_{HVF}$\tablenotemark{e}} &\colhead{R$_{HVF}^{Ca}$} \\
 \colhead{} &\colhead{(days)} &\colhead{(km s$^{-1}$)} &\colhead{(\AA)} &\colhead{(km s$^{-1}$)} &\colhead{(\AA)}}

\startdata
1991M & -7.8 & 13585 & 100 & 22817 & 124 & 1.24 \\
1991T & -7.5 & 10029 & 5 & 21636 & 8 & 1.64 \\
1994D & -11.1 & 14000 & 37 & 24000 & 192 & 5.16 \\
1994D & -9.5 & 12322 & 48 & 23177 & 75 & 1.58 \\
1994D & -8.5 & 11000 & 46 & 21200 & 39 & 0.83 \\
1997bq & -11.8 & 15500 & 59 & 24000 & 236 & 3.98 \\
1998dk & -7.1 & 11552 & 77 & 24000 & 207 & 2.68 \\
1998dm & -10.9 & 12179 & 69 & 22314 & 91 & 1.32 \\
\enddata
\tablenotetext{a}{Days since the B-band maximum light;}
\tablenotetext{b}{Velocity of the photospheric components of Ca~II~IR triplet;}
\tablenotetext{c}{Equivalent width (pEW) of the photospheric components of Ca~II~IR triplet;}
\tablenotetext{d}{Velocity of the HVF components of Ca~II~IR triplet;}
\tablenotetext{e}{Equivalent width (pEW) of the HVF components of Ca~II~IR triplet;}
\end{deluxetable}

\clearpage

\begin{deluxetable}{lccccccccc}
\tabletypesize{\scriptsize}\setlength{\tabcolsep}{0.02in}
\tablecaption{\label{T_sicam}Fit results of the Si~II~6355 and Ca~II~IR absorptions from Sample C} \tablewidth{0pt}
\tablehead{\colhead{SN} &\colhead{Phase\tablenotemark{a}} &\multicolumn{2}{c}{Si II $\lambda$ 6355} &\colhead{} &\multicolumn{5}{c}{Ca~II~IR}  \\
\cline{3-4} \cline{6-10} \\
 \colhead{} &\colhead{} &\colhead{$V^{Si}_{PHO}$\tablenotemark{a}} &\colhead{$pEW^{Si}_{PHO}$\tablenotemark{b}} &\colhead{} &\colhead{$V^{Ca}_{PHO}$\tablenotemark{c}} &\colhead{$pEW^{Ca}_{PHO}$\tablenotemark{d}} &\colhead{$V^{Ca}_{HVF}$}\tablenotemark{e} &\colhead{
$pEW^{Ca}_{HVF}$\tablenotemark{f}} &\colhead{R$_{HVF}^{Ca}$\tablenotemark{g}}
\\
 \colhead{} &\colhead{(days)} &\colhead{(km s$^{-1}$)} &\colhead{(\AA)} &\colhead{}
 &\colhead{(km s$^{-1}$)} &\colhead{(\AA)} &\colhead{(km s$^{-1}$)} &\colhead{(\AA)}}
\startdata
1989B & -1.0 & 10864 & 132 &  & 9876 & 125 & 16864 & 51 & 0.41 \\
1989M & -2.0 & 12496 & 133 &  & 12511 & 180 & 20211 & 36 & 0.20 \\
1991M & 1.2 & 14234 & 138 &  & 14351 & 166 & 23223 & 28 & 0.17 \\
1994S & 0.7 & 10278 & 85 &  & 10846 & 124 & 20011 & 16 & 0.13 \\
1996X & -2.1 & 10640 & 88 &  & 11176 & 122 & 21161 & 9 & 0.08 \\
1996X & -0.1 & 10671 & 94 &  & 11105 & 122 & 20440 & 16 & 0.13 \\
1996X & 0.9 & 10646 & 96 &  & 10822 & 128 & 18510 & 22 & 0.17 \\
1996ai & -2.1 & 10466 & 75 &  & 10445 & 91 & 19535 & 56 & 0.61 \\
\enddata

\tablecomments{}
\tablenotetext{a}{Velocity of the Si~II~6355;}
\tablenotetext{b}{Equivalent width of the Si~II~6355;}
\tablenotetext{c}{Velocity of the photospheric components of Ca~II~IR triplet;}
\tablenotetext{d}{Equivalent width of the photospheric components of Ca~II~IR triplet;}
\tablenotetext{e}{Velocity of the HVF components of Ca~II~IR triplet;}
\tablenotetext{f}{Equivalent width of the HVFs of Ca~II~IR triplet;}
\tablenotetext{g}{Ratio of line strength (EW) of HVF-Ca to that of PHO-Ca;}
\end{deluxetable}

\clearpage

\begin{table}
\centering\scriptsize \caption[htb]{\label{T_ks}\centering{P-values
of the K-S Tests for Different Samples of SNe Ia.}}
\begin{tabular}{|c|c|c|c|c|c|c|c|c|c|c|c|c|}
  \hline
    \tabincell{c}{Total Sample}                    &   \tabincell{c}{Subsample 1\\(number)}           & \tabincell{c}{Subsample 2\\(number)}           & \tabincell{c}{p-KS\\($\Delta$m$_{15}$(B))} &  \tabincell{c}{p-KS\\(B$_{max}$$-$V$_{max}$)} & \tabincell{c}{p-KS\\($V^{Si}_{max}$)}     &  \tabincell{c}{p-KS\\(Gal-type)}     & \tabincell{c}{p-KS\\($r_{SN}^{a}$)}  & \tabincell{c}{p-KS\\(M$^{Gal}_{K}$)} \\\hline
    \tabincell{c}{A\\(Si~II, $t<$-7d)}       & \tabincell{c}{R$^{Si}_{HVF}\geq0.2$ (29)} & \tabincell{c}{R$^{Si}_{HVF}<0.1$ (64)} & 0.035 & 0.129 & 0.002 & 0.476 & 0.317 & 0.657\\\hline
    \tabincell{c}{B$^b$\\Ca~II ($t<$-7d)}   & \tabincell{c}{R$^{Ca}_{HVF}\geq2.5$ (15)} & \tabincell{c}{R$^{Ca}_{HVF}<1.4$ (24)} & 0.001 & 0.351 & 0.073 & 0.260 & 0.581 & 0.461\\\hline
    \tabincell{c}{C\\Ca~II ($|t|<$3d)}      & \tabincell{c}{R$^{Ca}_{HVF}\geq0.8$ (22)} & \tabincell{c}{R$^{Ca}_{HVF}<0.5$ (98)} & 0 & 0.322 & 0.002 & 0.168 & 0.115 & 0.591\\\hline
    \tabincell{c}{A$^c$\\Si~II ($t<$-7d)}   & \tabincell{c}{R$^{Ca}_{HVF}\geq0.2$,\\$V^{Si}_{max}\geq12000$ (15)}  & \tabincell{c}{R$^{Si}_{HVF}\geq0.2$,\\$V^{Si}_{max}<12000$ (14)} & 0.026 & 0 & 0 & 0.371 & 0.002 & 0.658\\\hline
    \tabincell{c}{B$^d$\\ Ca~II ($t<$-7d)}   & \tabincell{c}{R$^{Ca}_{HVF}\geq2.5$,\\$V^{Si}_{max}\geq12000$ (7)}  & \tabincell{c}{R$^{Ca}_{HVF}\geq2.5$,\\$V^{Si}_{max}<12000$ (8)} & 0.214 & 0.014 & 0 & 0.504 & 0.064 & 0.805\\\hline
\end{tabular}
\flushleft{
$^a$ $r_{SN}=R_{SN}/R_{gal}$ is the projected galactocentric distance of the SN in its host galaxy, in units of the galaxy radius $R_{gal}$.
$^b$ subsample drawn from sample A;
$^c$ subsample drawn from sample A;
$^d$ subsample drawn from sample B;
}
\end{table}

\end{document}